\documentclass{emulateapj}       
\usepackage{amsmath}    
\usepackage{lscape}
\usepackage{graphics}

\newcommand{\cmm}{cm$^{-2}$}
\newcommand{\ecs}{erg\ cm$^{-2}$\ s$^{-1}$}
\newcommand{\ers}{erg\ s$^{-1}$}
\newcommand{\etal}{et al.}
\newcommand{\fx}{$S_{\rm X}$}

\newcommand{\lx}{$L_{\rm X}$}
\newcommand{\nh}{$N_{\rm H}$}
\newcommand{\lxz}{$L_{\rm X}$--$z$}

\newcommand{\na}{New A}

\journalinfo{Scheduled on The Astrophysical Journal, 635:???-???, 2005 December 20 }
\slugcomment{Received 2005 May 26; accepted 2005 August 15}

\shorttitle{HELLAS2XMM: THE HARD X--RAY LUMINOSITY FUNCTION OF AGN}
\shortauthors{LA FRANCA ET AL.}

\begin{document}

\title{The HELLAS2XMM survey. VII. The hard X--ray luminosity
function of AGN up to $z$=4: more absorbed AGN at low luminosities and high
redshifts }

\author{F. La Franca\altaffilmark{1}, 
F. Fiore\altaffilmark{2},
A. Comastri\altaffilmark{3},
G.C. Perola\altaffilmark{1},
N. Sacchi\altaffilmark{1},
M. Brusa\altaffilmark{4},
F. Cocchia\altaffilmark{2},
C. Feruglio\altaffilmark{2},
G. Matt\altaffilmark{1},
C. Vignali\altaffilmark{5},
N. Carangelo\altaffilmark{6},
P. Ciliegi\altaffilmark{3},
A. Lamastra\altaffilmark{1},
R. Maiolino\altaffilmark{7},
M. Mignoli\altaffilmark{3},
S. Molendi\altaffilmark{8},
S. Puccetti\altaffilmark{2}}

\altaffiltext{1}{Dipartimento di Fisica, Universit\`a degli Studi "Roma Tre",
Via della Vasca Navale 84, I-00146 Roma, Italy. \email{lafranca@fis.uniroma3.it}}
\altaffiltext{2}{INAF, Osservatorio Astronomico di Roma, Via , I-00100 Monteporzio, Italy}
\altaffiltext{3}{INAF, Osservatorio Astronomico di Bologna, Via Ranzani 1, I-40127 Bologna, Italy}
\altaffiltext{4}{Max Planck Institut f\"ur Extraterrestrische Phisik (MPE), Giessenbachstrasse, Postfach 1312,
85741 Garching, Germany}
\altaffiltext{5}{Dipartimento di Astronomia, Universit\`a di Bologna, Via Ranzani 1, I-40127 Bologna, Italy}
\altaffiltext{6}{Universit\`a di Milano-Bicocca, Piazza della Scienza 3, I-20126 Milano, Italy } 
\altaffiltext{7}{INAF, Osservatorio Astrofisico di Arcetri, Largo Fermi 5, I-50125 Firenze, Italy}
\altaffiltext{8}{INAF, IASF, Via Bassini 15, I-20133, Milano, Italy}

\begin{abstract}
  
  We have determined the cosmological evolution of the density of active
  galactic nuclei (AGN) and of their \nh\ distribution as a function of
  the un--absorbed 2--10 keV luminosity up to redshift 4.  We used the
  HELLAS2XMM sample combined with other published catalogs, yielding a
  total of 508 AGN. Our best fit is obtained with a luminosity-dependent density
  evolution (LDDE) model where low luminosity (\lx$\sim$10$^{43}$ erg
  s$^{-1}$) AGN peak at $z$$\sim$0.7, while high luminosity AGN
  (\lx$>$10$^{45}$ erg s$^{-1}$) peak at $z$$\sim$2.0.  A pure
  luminosity evolution model (PLE) can instead be rejected.
  
  There is evidence that the fraction of absorbed (\nh$>$10$^{22}$
  cm$^{-2}$) AGN decreases with the intrinsic X--ray luminosity, and
  increases with the redshift.
  
  Our best fit solution provides a good fit to the observed counts,
  the cosmic X--ray background, and to the observed fraction of
  absorbed AGN as a function of the flux in the
  $10^{-15}$$<$$S_{2-10}$$<$$10^{-10}$ \ecs range. We find that the
  absorbed, high luminosity (\lx$>$ 10$^{44}$ erg s$^{-1}$) AGN have a
  density of 267 deg$^{-2}$ at fluxes $S_{2-10}$$>$$10^{-15}$ \ecs.
  Using these results, we estimate a
  density of supermassive black holes in the local Universe of
  $\rho_{BH} = 3.2\ h^2_{70}\times 10^5$ M$_\odot$ Mpc$^{-3}$, which
  is consistent with the recent measurements of the black hole mass
  function in the local galaxies.

\end{abstract}


\section{Introduction}\label{intro}

The understanding of the history of accretion in the Universe and 
of the formation of massive black holes and their host galaxies relies on the
measurement of the active galactic nuclei (AGN) space density and
evolution.

According to the AGN unified model (Antonucci 1993) the viewing angle
between the observer and the symmetry axis of the nuclear structure is responsible for the
different classification.  In type 1 AGN the central engine is
directly visible.  Both the broad and narrow line emitting regions are
detected in the optical spectra along with a soft un-absorbed X--ray
spectrum. On the contrary, a type 2 AGN classification arises when the
broad line region and the soft X--rays are obscured by a dusty torus.

Until a few years ago the best measurements of the cosmological
evolution of the AGN luminosity function were essentially limited to
optically (e.g. La Franca \& Cristiani 1997, Croom \etal\ 2004), and
soft X--rays (e.g. Maccacaro et al.  1991, Miyaji \etal\ 2000) selected
type 1 AGN.  While there is evidence that type 2 AGN are about a
factor four more numerous than type 1 AGN (e.g.  Maiolino \& Rieke
1995; Risaliti et al. 1999), their relative space density beyond the local
Universe is basically unknown.  Assuming that the cosmological evolution
of type 1 and 2 AGN is the same, 
it was possible to simultaneously reproduce the
X--ray background spectrum and the X--ray counts (e.g. Setti \& Woltjer
1989; Comastri et al. 1995).  This simple picture was later slightly
modified in models where the fraction of type 2 AGN was assumed to
increase towards higher redshifts (e.g. Pompilio \etal\  2000, Gilli
\etal\  1999, 2001).  The selection of complete samples of type 2 AGN
is a difficult task. In the optical they are often
so dim that only the light of the host galaxy is visible; at 
$z >$1 even the latter has usually $R>$ 24.
In the soft X--rays bands even hydrogen column densities, \nh, of the order
of  10$^{21-22}$ cm$^{-2}$ may strongly suppress the flux.  In
the hard (2--10 keV) X--rays type 2 AGN selection is less biased
against, though the absorption due to large \nh\ column densities
(10$^{23-24}$ cm$^{-2}$) is not negligible especially at low
redshifts.

Early attempts to compute the hard X--ray luminosity function, based on
{\it ASCA} and {\it Beppo-SAX} observations (Boyle \etal , 1998; La
Franca \etal\ 2002 respectively) indicated a strong evolution for type
1 AGN, with a rate similar to that measured in the soft X--rays.
Unfortunately the low spatial resolution of the X--ray detectors
prevented an unambiguous identification of the type 2 AGN optical
counterparts, thus hampering a reliable determination of the type 2 AGN
space density.

Thanks to the high sensitivity and spatial resolution of the hard
X--ray detectors on board XMM-{\it Newton} and {\it Chandra}, it has
become possible to carry out AGN surveys less biased against X--ray
absorption and with more secure optical identifications.

However, already at fluxes fainter than S$_{2-10}$$\sim$10$^{-14}$
\ecs , a sizeable fraction of the X--ray sources have optical
magnitudes fainter than the spectroscopic limit of 8--10 meter class
optical telescopes, and thus the measure of their distance has to rely
on photometric redshifts, when it is not impossible altogether.

For these reasons, although the {\it Chandra} Deep Field North
  (CDF-N; Alexander \etal\ 2003) and the {\it Chandra} Deep Field
  South (CDF-S; Giacconi \etal\ 2002) surveys have resolved a fraction
  of the 2--10 keV XRB as large as 85--90\% (see also Brandt and
  Hasinger 2005), a clear picture of the AGN evolution able to
reproduce the whole set of observational constraints (i.e. soft and
hard X--ray counts, X--ray background, \nh\ and redshift distributions)
is still missing.

Attempts to take into account the redshift incompleteness
of X--ray selected AGN have been carried out by Cowie \etal\ (2003),
Fiore \etal\  (2003), Barger \etal\ (2005) combining data from deep
and shallow surveys.  They independently demonstrated that the AGN
number density for luminosities lower than $\sim$10$^{44}$ erg
s$^{-1}$ peaks at a lower redshift than that of high luminosity
objects.  Making use of an almost complete sample of 247 AGN from {\it
  Chandra}, {\it ASCA} and {\it HEAO1} surveys above a limiting flux
of S$_{2-10}$$>$$3.8\times 10^{-15}$ \ecs\ Ueda \etal\ (2003) were able
to estimate the hard X--ray luminosity function (HXLF) up to $z=3$.
They found that the fraction of the X--ray absorbed AGNs decreases
with the intrinsic luminosity and that the evolution of the AGN HXLF
is best described by a luminosity-dependent density evolution (LDDE).
Very similar results were also obtained by Hasinger \etal\ (2005)
using an almost complete sample of soft X--ray selected type 1 AGN.

In this paper we expand the study carried out by Fiore et al. (2003)
with the aim to compute the shape and evolution of the HXLF and \nh\ 
distribution of all AGN with \nh $<$10$^{25}$ cm$^{-2}$ up to $z\sim$
4. To reach such a goal it is necessary to cover the widest
possible range in the \lx-$z$-\nh\ space, and to take into account all
possible selection effects.  For these reasons we have used a large
AGN sample (about 500 objects) four times deeper than the Ueda et
al. (2003) sample. A new method to correct for the spectroscopic
incompleteness of faint X--ray sources is presented and discussed in
detail. The selection effects due to X--ray absorption are also
specifically discussed and estimated by an appropriate X--ray
``K--correction'' term.

%\medskip
The paper is structured as follows: in Section 2 we describe the
adopted X--ray samples; in Section 3 the method to compute the HXLF is
discussed. The results are presented in Section 4, discussed in
Section 5 and summarized in the last section.

Throughout this paper we call AGN all objects with an intrinsic
(corrected for \nh\ absorption) 2--10 keV X--ray luminosity
larger than 10$^{42}$ erg s$^{-1}$. In the last few years evidence
for a mismatch between optical (Type 1/2) and X--ray (un--absorbed/absorbed) 
classification has emerged (e.g. Fiore et al. 2000). In
this paper we refer to AGN1 if broad emission lines (rest frame
FWHM$>$2000 km s$^{-1}$) are present, while all remaining objects
(with or without narrow emission lines in the optical spectrum) are
called AGN2. If the rest frame column density is larger than
10$^{22}$ cm$^{-2}$ the AGN is classified as {\it absorbed}. The
adopted limit is well above the typical X--ray absorption by host
galaxy gas (disk, starburst regions, etc.) thus ensuring that the
measured column is most likely related to nuclear obscuration.  Unless
otherwise stated, all quoted errors are at the 68\% confidence level.
We assume $H_0 = 70$ km s$^{-1}$ Mpc$^{-1}$, $\Omega_m = 0.3$ and
$\Omega_{\Lambda} = 0.7$.

%%%%%%%%%%%%%%%%%%%%%%%%%%%%%%%%%%%%%%%%%%%%%%%%%%%%%%%%%%%%%%%%%%

%%%%%%%%%%%%%%%%%%%%%%%%%%%%%%%%%%%%%%%%%%%%%%%%%%%%%%%%%%%%%%%%%%
\bigskip
\section{Samples}

\begin{deluxetable}{lclll}
\tabletypesize{\scriptsize} 
\tablecaption{The samples}
\tablewidth{0pt} 
\tablehead{ \colhead{Sample } & \colhead{Flux limit
    } & \colhead{N$_S$ } & \colhead{N$_{sp}$ } &
    \colhead{$R_{lim}$ } \\
    \colhead{ (1)} & \colhead{ (2)} & \colhead{ (3)} & \colhead{ (4)}
    & \colhead{ (5)} } 
\startdata
  HEAO1 & 2.9$\times$10$^{-11}$ & \phn31 & \phn31 & \nodata \\
  AMSSn & 3.0$\times$10$^{-13}$ & \phn74 & \phn74 & \nodata \\
  HBS28 & 2.2$\times$10$^{-13}$ & \phn27 & \phn27 & \nodata \\
  H2XMM\tablenotemark{a} & 8.0$\times$10$^{-15}$ & 120 & 103 (\phn93) & 23.65 \\
  H2XMM\tablenotemark{b} & 8.0$\times$10$^{-15}$ & 110 &  \phn44  & 21.40 \\
  Lockman & 2.6$\times$10$^{-15}$ & \phn55 & \phn41 (\phn39) & 23.50 \\
  CDFN & 1.0$\times$10$^{-15}$ & 146 & 108 (102) & 24.65 \\
  CDFS & 1.0$\times$10$^{-15}$ & 127 & 102 (\phn98) & 25.00 \\
 \enddata 
\tablenotetext{a}{1df sample (Fiore \etal\ 2003).}
\tablenotetext{b}{0.5df sample (Cocchia \etal\ 2005).}
\tablecomments{
  In column (2) we give the flux limit of the samples in units erg
  cm$^{-2}$ s$^{-1}$. In column (3) ve give the total number of
  sources. In column (4) we give the number of sources brighter than
  the spectroscopic limit (in parenthesis those having redshift). In
  column (5) we give the spectroscopic completeness magnitude.  }
\label{tab_samp}
\end{deluxetable}

In order to cover the widest possible range of luminosities and
redshifts we combined the HELLAS2XMM sample (Fiore \etal\ 2003) with
other existing flux limited samples which allowed the estimates of the
rest frame \nh\ column density of each AGN.  Whenever possible, the
column density and the photon index ($\Gamma$) were determined with a
proper spectral analysis. Otherwise, we assumed $\Gamma=1.8$, and used
the hardness ratio to measure the $z$=$0$ column density ($N_{H0}$,
see also the discussion about the uncertainties of this approach
  in \S 4.1.1).  The rest frame column density (\nh ) was then
estimated by the relation $Log(N_H) = Log(N_{H0}) + 2.42Log(1+z)$,
which makes use of the Morrison \& McCammon (1983) cross sections,
including also the effects of the absorption edges, and assumes solar
abundances from Anders \& Grevesse (1989).

\begin{figure}
\epsscale{1.2}
%\plotone{Lx_z.eps}
\plotone{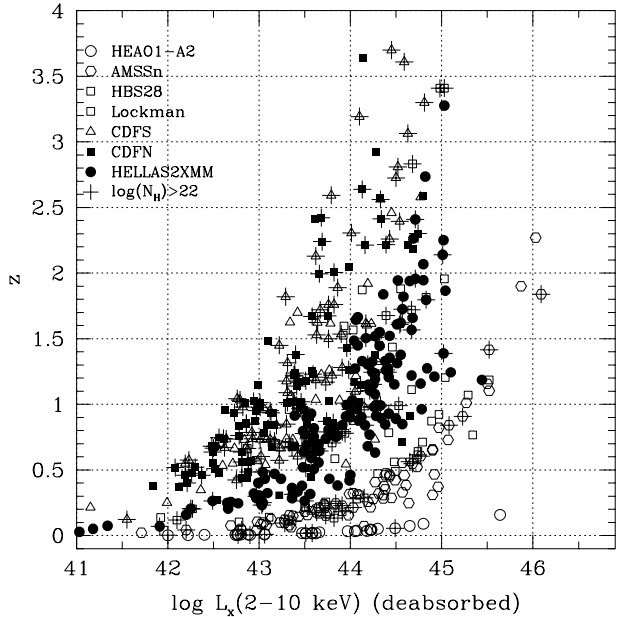}
\caption{
\lxz\ plane for all AGN used in this analysis. Different symbols corresponds to different surveys,
as labeled in the top left corner. Absorbed sources are
also highlighted by a cross.
 }\label{fig_lxz}
\end{figure}

\begin{figure}[t]
\epsscale{1.1}
%\plotone{PlotSamples.eps}
\plotone{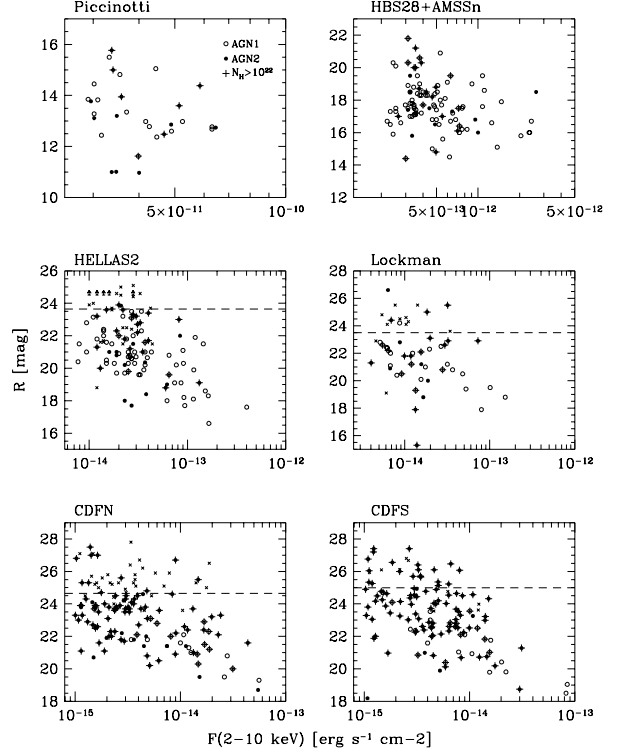}
\caption{
  R-band magnitude versus the 2--10 keV X--ray flux for all sources of
  the samples used in this analysis.  The dashed lines are the
  spectroscopic limits of completeness adopted in our analysis.  }
\label{fig_FxR}
\end{figure}

For those samples whose optical spectroscopic identifications are
incomplete, we chose the optical magnitude limit at which the samples
are almost spectroscopically complete. The incompleteness
is 6\% in the HELLAS2XMM,  Lockman, CDF-N and CDF-S samples). 
In these cases (as the X--ray--optical flux
distribution of the sources without redshift is almost similar to
that of the spectroscopically identified sources, and the fraction
of the unidentified sources is small) the sky coverage has been
reduced according to the fraction of spectra available.  
Table \ref{tab_samp} contains a
summary of the characteristics of each sample.  
The distribution in the \lx-$z$ space of all AGN from the
spectroscopically complete sub-samples used in our analysis are shown
in Figure \ref{fig_lxz}, while Figure \ref{fig_FxR} shows their
distribution in the $S_X$-R plane.

\bigskip
\subsection{The HELLAS2XMM sample}

We used the HELLAS2XMM 1df (1 degree field) sample (Fiore et al. 2003)
plus the recently available extension of 0.5 deg$^2$ (HELLAS2XMM 0.5df
Cocchia et al. 2005).  The HELLAS2XMM 1df sample contains 122 sources,
serendipitously detected in five XMM-{\it Newton} fields with \fx (2--10
keV)$>0.8\times 10^{-14}$ \ecs.  In our analysis we used the fluxes
and the column densities measured by X--ray spectral 
analysis (Perola et al. 2004). Among the 122 sources we discarded one
star (object n. 0537006) and one extended source (object n. 26900013).
For three sources with low signal-to-noise the hardness ratio
and redshift were used to estimate the rest frame \nh.  In summary, the
sample contains 120 sources, 115 optically identified, and 95 with
measured redshift and optically classified.  We restricted our
analysis to the sources brighter than R=23.65.  Down to this limit 93
out of 103 sources have been spectroscopically identified.

The HELLAS2XMM 0.5df sample consists of 110 objects brighter than
$S_X$(2--10 keV)$=8\times10^{-15}$ \ecs. Among them, 44 sources
brighter than R=21.4 (but otherwise randomly selected) have been
spectroscopically identified.

\subsection{The Piccinotti sample}

The Piccinotti sample is the brightest included in our analysis. It
has been obtained through observations carried out by the {\it HEAO1}
satellite, and contains 31 sources selected over an area of 26919
deg$^2$ down to $S_X$(2--10 keV)$=2.9\times10^{-11}$ \ecs (Piccinotti
et al. 1982).  The column densities have been taken from the literature,
and are derived from X--ray spectral analyses.

\bigskip
\medskip
\subsection{The AMSSn sample}

The AMSSn sample consists of 74 AGN at fluxes brighter than $S_X$(2--10
keV)$=3\times10^{-13}$ \ecs (Akiyama et al. 2003).  The total area
covered is 45 deg$^2$ at the fainter fluxes and rises up to $\sim$69
deg$^2$ at bright fluxes. The \nh\ column densities have been
  derived from the hardness ratios values.

\subsection{The HBS28 sample}

The HBS28 sample (Caccianiga \etal\ 2004) consists of 27 AGN and 1
star selected in the 4.5-7.5 keV band. The sources are brighter than
$S_X$(2--10 keV)$=2.2\times10^{-13}$ \ecs (assuming $\Gamma$=1.8) and
have been selected over 82 XMM-{\it Newton} pointed fields,
corresponding to a total of 9.756 deg$^2$.  All sources have been
spectroscopically identified, and their column densities have
  been measured through X--ray spectral fits.

\subsection{The Lockman Hole sample}

The Lockman Hole sample consists of 55 sources selected within
the 12 arcmin radius of the XMM-{\it Newton} observation.  The sources
are brighter than $S_X$(2--10 keV)$=2.6\times10^{-15}$ \ecs (Baldi
\etal\ 2002).  Optical identifications and X--ray spectral fits are
from Mainieri \etal\ (2002).  Spectroscopic redshifts and
classifications have been obtained for 41 objects, while 3 sources have
photometric redshifts.  We restricted our analysis to the sources
brighter than R=23.50. Down to this limit 39 out of 41 sources have
been spectroscopically identified.

\subsection{The CDF-N sample}

In order to reach almost spectroscopic completeness we
  have selected an X--ray bright subsample in the CDF-N.  The subsample 
consists of 146 sources (see Table 1) selected within the 10 arcmin
radius of the {\it Chandra} observation (Alexander \etal\ 2003). The
sky coverage reaches \fx(2--10 keV)$>10^{-15}$ \ecs\ in the inner 5.85
arcmin radius, \fx(2--10 keV)$>2.49\times10^{-15}$ \ecs\ in the annulus
between 5.85 and 6.5 arcmin radii, and \fx(2--10
keV)$>3.61\times10^{-15}$ \ecs\ in the annulus between 6.5 and 10.0
arcmin radii.  We used both spectroscopic and spectro-photometric
identifications and redshifts available from the literature (Barger
\etal\ 2003).  We restricted our analysis to sources brighter than
R=24.65. Down to this limit 102 out of 108 sources have been
spectroscopically identified. 
The \nh\ column densities have been
derived from the hardness ratios.

%\medskip
\subsection{The CDF-S sample}

Altough the CDF-S has been observed for 1 Ms instead of the 2 Ms
  spent in the CDF-N, we selected a spectroscopically complete X--ray bright
  subsample with the same sky coverage as for the
  CDF-N. Indeed, at our adopted flux limits, the difference in the
  exposure time does not affect the sky coverage.  The sample
consists of 127 sources (see Table 1; Giacconi et al. 2002 and
Alexander \etal\ 2003).  We used both spectroscopic and
spectro-photometric redshifts available from the literature (Szokoly
et al.  2004; Zheng \etal\ 2004). Moreover, given that both Szokoly
\etal\ (2004) and Zheng \etal\ (2004) identifications are based on the
X--ray source catalogue of Giacconi et al. (2002), we have revised some
optical/X--ray associations according to the improved astrometry
provided by Alexander \etal\ (2003). We restricted our analysis to
sources brighter than R=25.00. Down to this limit 98 out of 102
sources have been spectroscopically identified. The \nh\ column
  densities have been derived from the hardness ratios.
%\medskip

%%%%%%%%%%%%%%%%%%%%%%%%%%%%%%%%%%%%%%%%%%%%%%%%%%%%%%%%%%%%%%%%%%

\section{Method}

We searched for a functional fit to the density of the AGN as a
function of the un--absorbed 2--10 keV luminosity (\lx), the rest frame
absorbing column density (\nh), and the redshift ($z$). The method is
based on the comparison, through $\chi^2$ estimators, of the observed and
expected numbers of AGN (in the \lx-$z$ space) and of the \nh\ distributions,
obtained from computations which take into account all the
observational selection effects of the samples.

Once a HXLF evolution model is assumed, the number of expected
AGNs (E) in a given bin of the
\lx-$z$-\nh\ space is the result of the sum, over the number of
samples $N_{\rm samp}$, of the expected number of AGN in each sample
taking into account the area coverage of each $i$th sample
$\Omega_i(L,N_H,z)$, the \nh\ distribution $f(L_{\rm X},z;N_{\rm
H})$, and a completeness function $g(L_{\rm X},z,N_H,R_i)$, where
R$_i$ is the spectroscopic limit of completeness of the $i$th sample:

\begin{eqnarray}
\nonumber E=\sum_{i=1}^{N_{\rm samp}} \int \int \int \Phi (L_{\rm X}, z)f(L_{\rm X},z;N_{\rm H})\times\\
g(L_{\rm X},z,N_H,R_i)\Omega_i(L,N_H,z){dV\over{dz}} {\rm d Log} L_{\rm X}{\rm d}z{\rm d}N_H.
\end{eqnarray}

\subsection{The shape of the Luminosity Function}

In order to describe the evolution of the AGN, we used standard
functional forms, such as the pure luminosity evolution (PLE) model
and a luminosity-dependent density evolution (LDDE) model (see next
Section and, e.g., Boyle et al. 1998; Miyaji \etal\
2000; La Franca et al. 2002; Ueda et al.  2003). The HXLF,
representing the number density per unit comoving volume and per unit
Log \lx , as a function of \lx\ and $z$, was expressed as:

\begin{equation}
      \frac{{\rm d} \Phi (L_{\rm X}, z)}{{\rm d Log} L_{\rm X}}.
\end{equation}

We adopted a smoothly-connected two power-law form
to describe the present-day HXLF,

\begin{equation}
\frac{{\rm d} \Phi (L_{\rm X}, z=0)}{{\rm d Log} L_{\rm X}} 
= A [(L_{\rm X}/L_{*})^{\gamma 1} + (L_{\rm X}/L_{*})^{\gamma 2}]^{-1}.
\end{equation}

\subsection{The K-correction}

In order to convert the observed 2--10 keV fluxes (\fx ) to the
intrinsic 2--10 keV luminosities (\lx ) and vice-versa, for each
observed or ``expected'' AGN with a given \nh , a K-correction has
been computed by assuming a photon index $\Gamma=1.8$, an exponential
cutoff ($e^{-E/E_C}$) at $E_C=200$ keV, and the corresponding
photoelectric absorption (see \S 4.1.1 for a discussion on the use of
different K-corrections).

\subsection{The completeness function}

All the faint samples used in our analysis (HELLAS2XMM, Lockmann,
CDF-S, CDF-N) are nearly spectroscopically complete down to a certain
optical limit magnitude ($R$$=$$21.4-25$, see Table \ref{tab_samp}).
In order to compute the number of expected AGN in a certain bin of the
\lx-$z$-\nh\ space, we introduced the {\it completeness function}
$g(L_{\rm X},z,N_H,R)$ which provides the probability that a given AGN
with luminosity \lx , redshift $z$ and column density \nh , had an
apparent R-band magnitude brighter than the spectroscopic limits of
completeness R of each sample.

\begin{figure}[t]
\epsscale{1.15}
%\plotone{LxLo2.eps}
\plotone{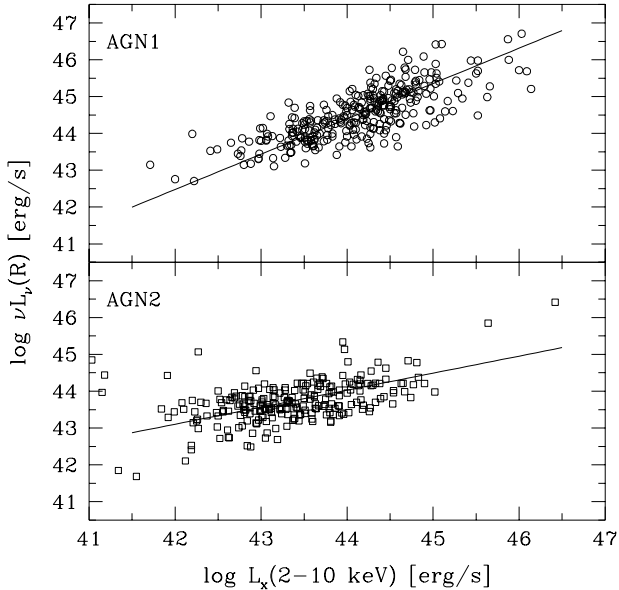}
\caption{Log$L_X$--Log$L_R$ relation for optical AGN1
  and AGN2. The continuous lines correspond to eq. \ref{eq_slpAGN1}
  and \ref{eq_slpAGN2}.}
\label{fig_lxlR}
\end{figure}

For this reason we derived an empirical relationship between the
un--absorbed X--ray luminosity \lx\ and the optical luminosity $ L_{\rm
  R}$\footnote{The $ L_{\rm R}$ luminosity is in erg s$^{-1}$ ($\nu L_{\nu}$),
  computed at 660 nm, where the flux is $f{\rm [erg\ s^{-1}\ cm^{-2}\ Hz^{-1}]}
  =2.84\times10^{-20}\times10^{-0.4R}$ (Zombeck 1990).}  for AGN1 and
AGN2, and measured their spread (see Figure \ref{fig_lxlR}).  For
AGN1 we found:
\begin{equation}
{\rm Log} L_{\rm R} = 0.959(\pm .025)\times {\rm Log} L_{\rm X} +2.2(\pm 1.1),
\label{eq_slpAGN1}
\end{equation}

\noindent
with a 1$\sigma$ dispersion of 0.48 (in Log$L_R$ units) around the
best fit solution. The linear correlation coefficient is r=0.773,
corresponding to a negligible ($<$10$^{-13}$) probability that the
data are consistent with the null hypothesis of zero correlation. For
AGN2 a flatter relation was found:
\begin{equation}
{\rm Log} L_{\rm R} = 0.462(\pm .026)\times {\rm Log} L_{\rm X}+23.7(\pm 1.1),
\label{eq_slpAGN2}
\end{equation}

\noindent
with a 1$\sigma$ dispersion of 0.40 (in Log$L_R$ units), and a linear
correlation coefficient r=0.462, again corresponding to a negligible ($<$
2$\times$10$^{-13}$) probability that the data are consistent with the
null hypothesis of zero correlation.  In order to compute the above
relationships a linear least squared method with errors (assumed 0.2
dex) in both axes has been used.  The difference between the two
relations should be attributed to the dominance in the optical of the
AGN component in the AGN1, which produces an almost linear
relationship between X--ray and optical luminosity (see La Franca et
al., 1995 for similar results in the soft X--rays).  In
AGN2, where the nucleus is obscured, the optical luminosity is
instead dominated by the host galaxy (see also Fiore \etal\ 2003).

For each pair of un-absorbed X--ray luminosity and redshift, the above
relationships ({\it with their spreads}) can be used to compute the
probability of an AGN to appear brighter than a certain optical
magnitude, and thus be spectroscopically identified.  The observed
spreads of the two relationships are due to a combination of the
intrinsic spread with the observational uncertainties. Given our aims, 
both effects should be taken into
account, and we have thus not subtracted the contribution of the
observational uncertainties from the spread estimates.  To
choose which \lx -$L_R$ relationship to use (eq.
\ref{eq_slpAGN1} or eq. \ref{eq_slpAGN2}), we need also to know the
probability of an AGN to appear as an AGN1 (or, its complement, an AGN2)
as a function of \lx, \nh\ and $z$: Q1(\lx, $z$, \nh).  This
probability was estimated from the sample itself as described below.

\begin{figure}
\epsscale{1.15}
%\plotone{SM_PlotLxNH0_2.eps}
\plotone{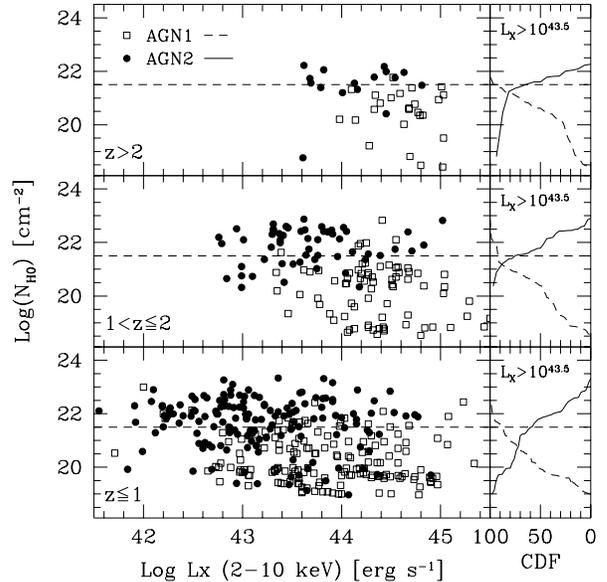}
\caption{
  Log\lx --Log$N_{H0}$ plane at $z\leq 1$ ($\it bottom$), $1<z\leq 2$
  ($\it middle$), $z>2$ ($\it up$).  Open squares are optical AGN1,
  filled circles are optical AGN2.  On the right side the dashed lines
  are the cumulative distribution functions of the N$_{H0}$ values for
  AGN1 with \lx$>$10$^{43.5}$ \ers , while the continuous lines are the complement of the
  cumulative distribution function of the N$_{H0}$ values for AGN2 with \lx$>$10$^{43.5}$ \ers . }
\label{fig_lxNh}
\end{figure}

\begin{figure}[t]
\centering
\resizebox{\hsize}{!}
{\includegraphics[angle=-90]{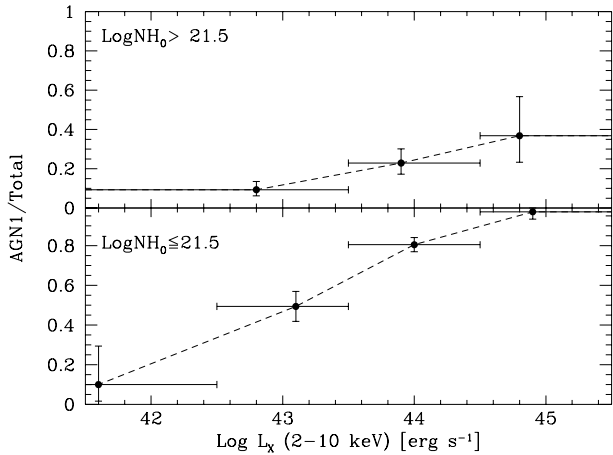}}
\caption{
  Fraction of optical type 1 AGN with N$_{H0}$$>$10$^{21.5}$ \cmm\
  ($up$), and N$_{H0}$$\leq$10$^{ 21.5}$ \cmm\ ($bottom$) as function
  of the intrinsic luminosity \lx.}
\label{fig_frAGN1}
\end{figure}

Figure \ref{fig_lxNh} shows the distribution of the observer frame
column density N$_{H0}$ as a function of \lx\ for AGN1 and AGN2, in
three redshift intervals.  Here we do not use the rest
frame \nh, but instead the observer frame N$_{H0}$ which is equivalent
to an hardness ratio (see also Hasinger 2003).  As can be seen 
in Figure \ref{fig_frAGN1}, the probability to find an AGN1 is not
only dependent on N$_{H0}$, but depends also on the luminosity.
The probability to find an AGN1 increases with increasing
luminosities, and there is a relevant fraction of low luminosity
(\lx$<$10$^{43}$ \ers) un--absorbed objects which are AGN2, while a
fraction of the high luminosity (\lx$>$10$^{45}$ \ers) absorbed
objects are AGN1.  This result, if it is not due to the contamination
by the galaxy light in the lower luminosity AGN2, is against the
simplest version of the AGN unified model. The analysis of this issue
is beyond the scope of this paper (see Panessa \& Bassani 2002, Page
\etal\ 2003, Steffen \etal\ 2003, Ueda \etal\ 2003, Brusa \etal\ 2003,
Perola \etal\ 2004 and Barger \etal\ 2005 for similar results and
discussions. See also \S 4.6)

As Figure \ref{fig_lxNh} shows, there is no
evidence of a dependence on redshift of the distribution of AGN1 and AGN2 as a
function of \lx\ and N$_{H0}$. We have thus
estimated the probability of an AGN to appear as an AGN1 as a function
of \lx\ and N$_{H0}$ only, by assuming no dependence on redshift.
This probability has been estimated as a function of \lx\, in two bins
of N$_{H0}$\footnote{ We chose to use, here, the observed column
  densities (N$_{H0}$) instead of the intrinsic ones (\nh ) in order
  to eliminate the dependencies on the redshift.  A constant (with
  $z$) N$_{H0}$$=$10$^{21.5}$ \cmm\ separation limit corresponds to a
  shifts of this limit towards higher values of \nh\ with increasing
  redshift (as the intrinsic and the $z$$=$0 
  column densities are related by the equation $log(N_H) = log(N_{H0})
  + 2.42log(1+z)$).  We will come back to this point in the next
  Sections. However, we wish to stress here that the above
  relationships have been derived only in order to correct the samples
  for spectroscopic incompleteness, and that N$_{H0}$=10$^{21.5}$
  \cmm\ should not be meant as the working separation limit between
  absorbed and un--absorbed AGN, which (as defined in \S 1) instead is
  N$_{H}$=10$^{22}$ \cmm.}: at N$_{H0}$$\leq$10$^{21.5}$ \cmm\ and
N$_{H0}$$>$10$^{21.5}$ \cmm . The values of the probability of an AGN
to appear as an AGN1 in these two N$_{H0}$ intervals are shown in
Figure \ref{fig_frAGN1}.

%{\includegraphics[angle=-90]{SM_FracAGN1_2.eps}}

We caution the reader that, due to inhomogeneities on the quality of
the spectroscopic classification of the samples used, the above
measure of the fraction of AGN1 as a function of \lx\ and N$_{H0}$ has
uncertainties that are difficult to quantify.  However, these
estimates are only used to derive which fractions of the unidentified
AGN will follow the two \lx -$L_{\rm R}$ relationships shown in eq.
\ref{eq_slpAGN1} and \ref{eq_slpAGN2}.  The absence of many outliers
in the \lx - $L_{\rm R}$ relationships for AGN1 and AGN2 shown in
Figure \ref{fig_lxlR}, demonstrates qualitatively that classification
errors should not be very large. This, in principle, does not imply
that more accurate spectroscopy would not change the optical
classification of the AGN, but that the spectroscopy is accurate
enough, for our purposes, to decide which of the two \lx - $L_{\rm R}$
relationships the AGN would follow.  However, the completeness
correction is computed under the assumption that the measured fraction
of AGN1 as a function of \lx\ and N$_{H0}$ and the derived two \lx -
$L_{\rm R}$ relationships for AGN1 and AGN2 hold also for the higher
redshift, optically fainter, unidentified population.  We will discuss
in \S 4 how much the uncertainties on these assumptions would affect
our results.

In summary, the completeness function $g(L_X,z,N_H,R_i)$ was computed
as follows: for each given AGN having intrinsic luminosity \lx ,
redshift $z$, and absorption column-density \nh , a) the N$_{H0}$ was
derived according to the equation $Log(N_{H0}) = Log(N_H) -
2.42Log(1+z)$, b) the probabilities to be an AGN1 (Q1) and AGN2
(1-Q1) were estimated according to the values plotted in Figure
\ref{fig_frAGN1}, and then c) according to eq. \ref{eq_slpAGN1} and
\ref{eq_slpAGN2} {\it and their spreads}, the two probabilities (for
the fraction of AGN1 and AGN2) to be brighter than the
spectroscopic limit R$_i$ of the $i$th sample were computed and
summed.

\subsection{The \nh\ function}

To describe the distribution of the spectral parameters of the
AGNs at a given
luminosity and redshift, we introduced the \nh\ function, $ f (L_{\rm
X}, z; N_{\rm H}), $ a probability-distribution function for the
absorption column-density as a function of \lx\ and $z$. The \nh\
function (in Log\nh$^{-1}$ units) is normalized to unity
at each redshift,  over the \nh\ interval 20$<$Log\nh$<$25:

\begin{equation}
\int_{20}^{25} f (L_{\rm X}, z; N_{\rm H}) {\rm d Log} N_{\rm H} = 1.
\label{eq_intnh}
\end{equation}

The objects have been grouped into 5 bins of \nh , $\Delta$Log\nh =1
wide, and centered at Log\nh= 20.5, 21.5, 22.5, 23.5, 24.5. The first
bin includes all the AGNs having \nh$<$10$^{21}$ \cmm.

%{\includegraphics[angle=-90]{Fig_Frac_FLAT.eps}}
\begin{figure}
\centering
\resizebox{\hsize}{!}
{\includegraphics[angle=-90]{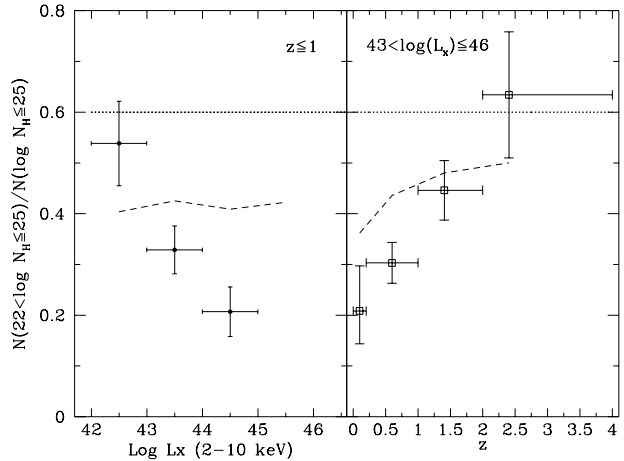}}
\caption{
  Observed fraction of absorbed (\nh $>$10$^{22}$ \cmm) AGN as a
  function of \lx\ (left panel) and $z$ (right panel).  The dotted
  lines are the intrinsic fractions assuming a constant flat \nh\ 
  distribution in the range 10$^{20}$$<$\nh$<$10$^{25}$ \cmm.  The
  dashed lines are the expectations taking into account the selection
  effects.  }
\label{fig_NHlzLDDEs}
\end{figure}

%{\includegraphics[]{Fig_HistNH_FLAT.eps}}
\begin{figure}
\epsscale{1.15}
\plotone{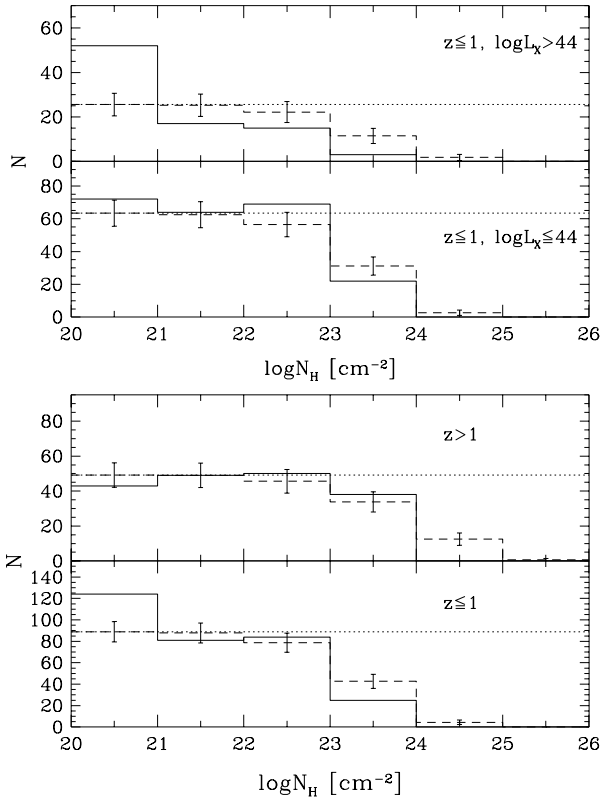}
\caption{
  \nh\ distributions in various luminosity and redshift intervals.  {\it
    Top}. High luminosity and low luminosity AGN at $z$$<$1.  {\it
    Bottom}. High and low redshift AGN.  The continuous lines are the
  observed distributions, the dotted lines are the assumed (constant
  flat) \nh\ distributions, while the dashed lines are the
  expectations taking into account the selection effects.  }
\label{fig_NHLDDEs}
\end{figure}

In Figure \ref{fig_NHlzLDDEs} the observed fraction of absorbed
(\nh$>$10$^{22}$ \cmm) AGN as a function of \lx\ and $z$ is shown. The
dotted lines correspond to the fraction of absorbed objects if a flat
\nh\ distribution in the range 10$^{20}$$<$N$_H$$\leq$10$^{25}$ \cmm\ 
were assumed, with no selection effects taken into account.
The dashed lines show our predictions when these
effects are included.  Such a model does not
provide a good fit to the data points, where a decrease with the
intrinsic luminosity and an increase with the redshift is observed.
This behavior is also evident in Figure \ref{fig_NHLDDEs}. Bearing in
mind that the \nh\ estimates are affected by uncertainties that can be
as large as one decade, from the analysis of Figure \ref{fig_NHLDDEs}
it appears that the assumption of a flat \nh\ distribution
produces an expected distribution roughly in agreement with the observed
one at \nh$>$10$^{21}$ \cmm.  Hence, the observed change of the
fraction of absorbed AGN as a function of \lx\ and $z$ (see Figure
\ref{fig_NHlzLDDEs}), could be mainly attributed to a change of the
fraction of AGN with \nh$<$10$^{21}$ \cmm. This will be our working
hypothesis, which we will analyze in the next Sections.

We have thus assumed a flat \nh\ distribution between \nh = 10$^{21}$
\cmm\ and \nh=10$^{25}$ \cmm, while allowing the fraction of objects
with \nh$<$10$^{21}$ \cmm to vary.  We introduced a linear dependence
of the fraction of objects with \nh$<$10$^{21}$ \cmm\ ($\Psi = f
(L_{\rm X}, z; LogN_H<21)$) on both Log\lx\ and $z$:

\begin{equation}
\Psi(L_{\rm X},z) = \psi[(LogL_{\rm X} -44)\beta_L+1][(z-0.5)\beta_z+1],
\label{eq_nh}
\end{equation}

\noindent
where $\psi$ is the fraction of objects with
\nh$<$10$^{21}$ \cmm\ at \lx = 10$^{44}$ \ers\ and $z$=0.5, and
$\beta_L$ and $\beta_z$ are the slopes of the linear dependences on
\lx\ and $z$ respectively.  This choice is the simplest possible
according to the quality of the data.  The function holds for the 
ranges $0.25\leq z\leq 2.75$ and
10$^{42.5}$$\leq$\lx$\leq$10$^{45.5}$ \ers.  At redshifts and
luminosities outside these ranges the fraction was kept constant,
equal to the values assumed at the limits of the ranges. Obviously
$\Psi$ could take all values in the range [0,1].  This corresponds
to an allowed fraction of absorbed objects (\nh$>$10$^{22}$ \cmm ) in
the range 0\% - 75\%. Indeed, according to eq. \ref{eq_intnh} and the
assumption of a flat \nh\ distribution at \nh$>$10$^{21}$ \cmm, the
fraction of absorbed AGN turns out to be:

\begin{equation}
{{N(22<LogN_H\leq25)}\over{N(LogN_H\leq25)}}= 0.75(1-\Psi).
\end{equation}

As clear from Figure \ref{fig_NHLDDEs}, no object with
\nh$>$10$^{25}$ \cmm\ is either expected or observed. Thus, we limited
our statistical analysis of the evolution of the AGN to the objects
having \nh$\leq$10$^{25}$ \cmm (see eq. \ref{eq_intnh}).  However,
when we will predict the number counts, X--ray background and the
accretion history (\S 4.3, \S 4.4 and \S 5.2), we will include in the \nh\
distribution a number of objects with 10$^{25}$$<$\nh
$\leq$10$^{26}$ \cmm , equal to that in the interval 10$^{24}$$<$\nh
$\leq$10$^{25}$.

%\bigskip
\subsection{$\chi^2$ fitting}

In order to find the best fitting model we choose two $\chi^2$
estimators as figure of merit functions.

The first estimator ($\chi^2_{\rm LF}$) is related to the shape and
evolution of the HXLF and is obtained by comparing the expected and
observed numbers of AGN in 24 bins, covering the whole
sampled Hubble space (\lxz ; see Figure \ref{fig_xlfLDDEslz} for an
example of the binning).  Computations have been carried out
in the 0$<$$z$$<$4.5 redshift range, and in the
10$^{42}$$<$\lx$<$10$^{47}$ \ers\ luminosity range. A total of 508
  AGN were used. 190  had the \nh\ column densities directly
  measured from X--ray spectroscopic analysis.

The second estimator ($\chi^2_{\rm N_H}$) is related to the \nh\
function, $ f (L_{\rm X}, z; N_{\rm H}) $, i.e. the shape of the \nh\
distribution and its dependence on \lx\ and $z$. One contingency table
was created dividing the objects with column densities higher or
lower than \nh = 10$^{22}$ cm$^{-2}$ into 5 further bins in the \lxz\
space. The $\chi^2_{\rm N_H}$ estimator was computed by comparing the
expected and observed number of AGN in the total 10 ( 2$\times$5 )
bins.

The reasons for using two different $\chi^2$ estimators are:
a) the number of objects is too
small to construct a single $\chi^2$ estimator using bins in
the three-dimensional space \lx\--$z$--\nh\ ; b) the two $\chi^2$
estimators cannot be summed as the data used are not independent.
The shape and the evolution of the HXLF is only marginally dependent
on the shape and evolution of the \nh\ distribution (we checked that
the best fit parameters of the HXLF vary within the 1$\sigma$
uncertainties when the parameters of the \nh\ distribution are left to
vary within a 3$\sigma$ range of their best-fit values).

The final fit was obtained by iteratively searching for the lowest
values of $\chi^2_{\rm LF}$ and $\chi^2_{\rm N_H}$ in turn, until the
changes on the two $\chi^2$ estimators were smaller then
0.1\footnote{This is a small enough interval, as the variance on the
  $\chi^2$ estimator is 2N$_d$, where N$_d$ are the degrees of
  freedom.  Variations of $\Delta\chi^2$=0.1 correspond to confidence
  levels of less than 2\% and 3\% for $\chi^2_{\rm LF}$ and
  $\chi^2_{\rm N_H}$ respectively.}.  For each model the probabilities
for $\chi^2_{\rm LF}$ and $\chi^2_{\rm N_H}$, according to the
corresponding degrees of freedom, were computed. Confidence regions of
each parameter have been obtained by computing $\Delta\chi^2$ at a
number of values around the best fit solution, while letting the other
parameters free to float (see Lampton \etal\ 1976).  The 68\%
confidence regions quoted correspond to $\Delta\chi^2$=1.0.  Moreover,
in order to use an un-binned goodness of fit test of the HXLF models,
we used also a bi-dimensional Kolmogorov-Smirnov test (2D--KS, Fasano
\& Franceschini 1987) on the Hubble (\lxz ) space.

%%%%%%%%%%%%%%%%%%%%%%%%%%%%%%%%%%%%%%%%%%%%%%%%%%%%%%%%%%%%%%%%%%
\section{Results}

\begin{table*}[t]
\caption{2-10 keV AGN LF parameters}
%17+++++++++++++123456789012345678901234567890
\begin{tabular}{llccccccccccccccccc}
\hline
\hline
\footnotesize
\smallskip
 Model &
  A\tablenotemark{a}&
 $p1$ &
 $p2$ &
 $z_{cut}$ &
  $\alpha$ &
 Log$L_a$ \tablenotemark{b}&
 Log$L^*$ \tablenotemark{b}&
 $\gamma_1$ &
 $\gamma_2$ &
 $\psi$ &
 $\beta_L$ &
 $\beta_z$ &
 P$_{_{LF}}(\chi^2)$\tablenotemark{c} &
 P$_{_{LF}}(KS)$\tablenotemark{c}  &
 P$_{N_H}(\chi^2)$\tablenotemark{c} &
 XRB$_{2-10}$\tablenotemark{d} \\
%\hline

\tableline
\\
\multicolumn{17}{c}{LDDE}\\
\tableline
 1     & 1.48& 4.37 & -1.19 & 2.39 & 0.20 & 45.74 &  44.26 & 0.94 & 2.35 & 0.26 & 0.00 & -0.00 & 20 & \phn6 & \phn0.07 &1.76\\
 2     & 1.50& 4.39 & -1.14 & 2.41 & 0.20 & 45.74 &  44.25 & 0.97 & 2.36 & 0.29 & 0.43 & -0.00 & 20 & \phn6 & \phn0.5\phn  &1.78\\
 3     & 1.39& 4.48 & -1.19 & 2.39 & 0.20 & 45.74 &  44.26 & 0.94 & 2.35 & 0.30 & 0.00 & -0.33 & 19 & \phn5 & \phn0.09 &1.75\\
 4     & 1.21& 4.62 & -1.15 & 2.49 & 0.20 & 45.74 &  44.25 & 1.01 & 2.38 & 0.44 & 0.62 & -0.51 & 20 & \phn7 & 83\phm{.00}   &1.81\\
 5 \tablenotemark{e}  & 1.29& 4.85 & -1.03 & 2.45 & 0.22 & 45.73 &  44.23 & 1.09 & 2.44 & 0.36 & 0.67 & -0.00 & 33 &  21 & 20\phm{.00}   &2.16\\
\\
 Errors & 5\%  &$^{+0.26}_{-0.26}$&$^{+0.72}_{-0.68}$&$^{+0.82}_{-0.68}$&$^{+0.04}_{-0.03}$&$^{+0.58}_{-0.63}$&$^{+0.18}_{-0.18}$&$^{+0.08}_{-0.10}$
&$^{+0.13}_{-0.11}$&$^{+0.04}_{-0.05}$&$^{+0.14}_{-0.13}$ &$^{+0.14}_{-0.17}$ \\
\tableline
\\
\multicolumn{17}{c}{LDDE}\\
\tableline
 6     & 6.18& 3.22 & \nodata   & 1.08 & \nodata  & \nodata  &  43.79 & 0.95 & 2.74 & 0.46 & 0.64 & -0.58 &~~9 & ~~3 & 63~~~~~~   &2.54\\
\\
 Errors & 5\%  &$^{+0.13}_{-0.26}$& \nodata &$^{+0.08}_{-0.06}$& \nodata & \nodata &$^{+0.15}_{-0.11}$&$^{+0.06}_{-0.07}$
&$^{+0.27}_{-0.23}$&$^{+0.04}_{-0.05}$&$^{+0.14}_{-0.13}$ &$^{+0.14}_{-0.17}$ \\
\tableline

\end{tabular}
\tablenotetext{a}{In units of 10$^{-6}$ $h_{70}^3$ Mpc$^{-3}$. }
\tablenotetext{b}{In units of \ers.}
\tablenotetext{c}{Probability values in \% units.}
\tablenotetext{d}{In units 10$^{-11}$ erg cm$^{-2}$ s$^{-1}$ deg$^{-2}$.}
\tablenotetext{e}{Only Piccinotti, AMSSn and CDF-N samples used.}
\label{tab_fit}
\end{table*}
\noindent

\subsection{The LDDE model}

By introducing the evolution factor

\begin{equation}
e(z) = \left\{ \begin{array}{ll}
   (1+z)^{p1}  &(z \leq z_{\rm c})\\
    e(z_{\rm c})[(1+z)/(1+z_{\rm c})]^{p2}  &(z > z_{\rm c}),
\end{array}
\right. \
\label{eq-evolv}
\end{equation}

\noindent
the pure density evolution (PDE) model is expressed as
\begin{equation}
\frac{{\rm d} \Phi (L_{\rm X}, z)}{{\rm d Log} L_{\rm X}} 
= \frac{{\rm d} \Phi (L_{\rm X}, 0)}{{\rm d Log} L_{\rm X}} e(z).
\label{eq-PDE}
\end{equation}

The $z_{\rm c}$ parameter represents the redshift at which
the evolution stops. $p1$ is the
parameter characterizing the rate of the evolution, while $p2$ is
usually negative and characterizes the rate of the counter-evolution
of the HXLF at $z$$>$$z_c$.

The LDDE model is obtained by introducing in the PDE model a
luminosity dependence of $z_{\rm c}$, assumed to be a power law:

\begin{equation}
z_{\rm c}(L_{\rm X}) = \left\{ \begin{array}{ll}
 z_{\rm c}^* & (L_{\rm X} \geq L_a) \\
z_{\rm c}^* (L_{\rm X}/L_a)^\alpha & (L_{\rm X}<L_a).
\end{array}
\right. \
\end{equation}

%{\includegraphics[]{xlfLDDEslz.eps}}
\begin{figure}[b]
\centering
\resizebox{\hsize}{!}
{\includegraphics[]{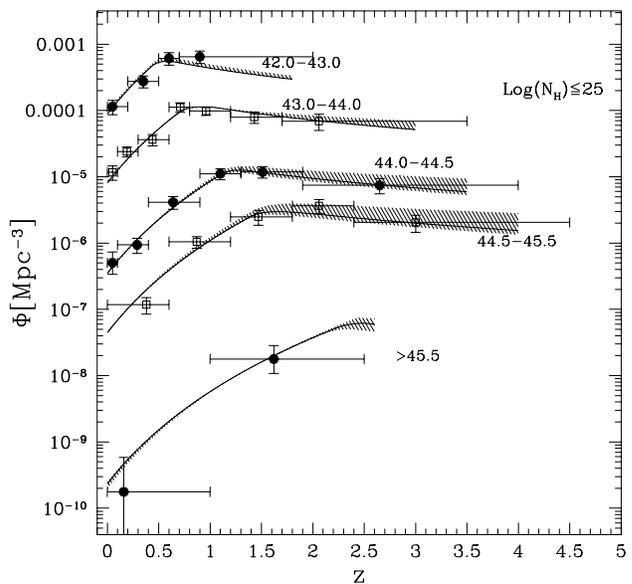}}
\caption{
  Density of AGN in luminosity bins as a function of redshift. The
  continuous lines are the best fit values of the LDDE model with an
  evolving \nh\ distribution depending on \lx\ and $z$ (fit \#4 in
  Table \ref{tab_fit}).  Data have been plotted using the ``$N^{\rm
    obs}/N^{\rm mdl}$ method'' (see \S4.1). The dashed areas
    are the largest allowed regions due to uncertainties in the
    completeness correction method used (see \S 4.1.1).
}
\label{fig_xlfLDDEslz}
\end{figure}

The above parameterization has been introduced by Ueda et al. (2003)
in order to allow for a change with luminosity of the redshift at
which the density of AGNs peaks (see also Miyaji \etal\ 2000 for a
similar LDDE parameterization).  This behavior is also apparent in our
data (see, e.g.  Figure \ref{fig_xlfLDDEslz}).

In order to plot the HXLF we adopted the ``$N^{\rm obs}/N^{\rm mdl}$
method'' (La Franca \& Cristiani 1997), where the best-fit model
multiplied by the ratio between the number of observed sources and
that of the model prediction in each \lxz\ bin is plotted. Although
model dependent (especially when large bins are used), this technique
is the most free from possible biases, compared with other methods
such as the conventional $1/V_a$ method.  The attached errors are
estimated from Poissonian fluctuations (1$\sigma$) in the observed number of
sources according to the Gehrels (1986) formulae.

\begin{figure}
\centering
\resizebox{\hsize}{!}
{\includegraphics[]{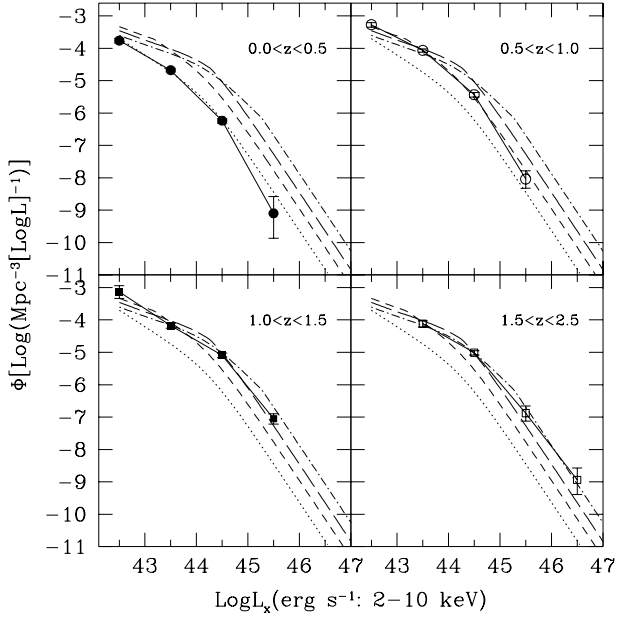}}
\caption{
  Density of AGN as a function of luminosity in four redshift
  intervals. The values are plotted at the central redshift of the
  intervals.  The dashed lines are the best fit densities of the LDDE
  model with an evolving \nh\ distribution depending on \lx\ and $z$
  (fit \#4 in Table \ref{tab_fit}).  Data have been plotted using the
  ``$N^{\rm obs}/N^{\rm mdl}$ method'' (see \S4.1).  }
\label{fig_lfLDDEslz}
\end{figure}

We simultaneously fitted the parameters of the HXLF and of the possible
dependencies of the \nh\ distribution on \lx\ and $z$. As shown in
Table \ref{tab_fit}, the LDDE model provides a good fit to the data
regardless of the adopted \nh\ distribution (see Figures
\ref{fig_xlfLDDEslz} and \ref{fig_lfLDDEslz}).

%{\includegraphics[]{LFLDDEslz.eps}}

\begin{figure}
\epsscale{1.15}
\plotone{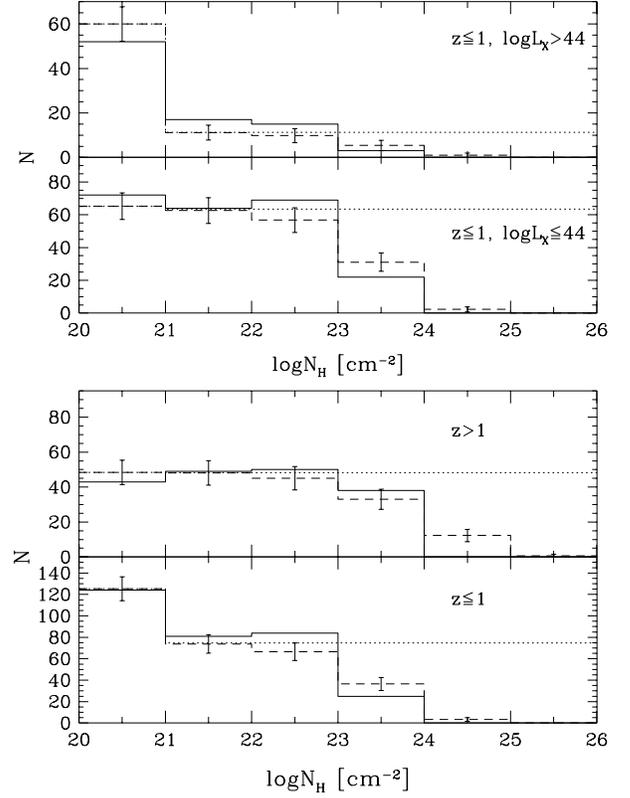}
\caption{
  \nh\ distributions in various luminosity and redshift intervals. {\it
    Top}. High luminosity and low luminosity AGN at $z$$<$1.  {\it
    Bottom}. High and low redshift AGN. The continuous lines are the
  observed distributions, the dotted lines are the assumed \nh\ 
  distributions (fit \#4 in Table \ref{tab_fit}: evolving \nh\ 
  distribution with a LDDE HXLF evolution), while the dashed lines are
  the expectations taking into account the selection effects.  }
\label{fig_NHLDDEslz}
\end{figure}

\begin{figure}
\centering
\resizebox{\hsize}{!}
{\includegraphics[angle=-90]{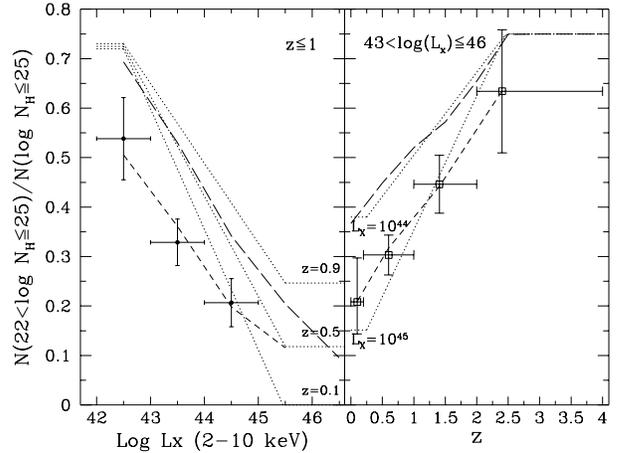}}
\caption{
Observed fraction of absorbed (\nh$>$10$^{22}$ \cmm) AGN as a function of
\lx\ and $z$. The dotted lines are examples of the intrinsic assumed distributions
at various luminosities and redshifts (LDDE model, fit \#4 in Table
\ref{tab_fit}).  The long dashed lines are the corresponding average
intrinsic assumed distributions of the sample used.  The short dashed
lines are the expectations taking into account the selection effects.
}
\label{fig_FracLDDEslz}
\end{figure}

According to these fits, the redshift of the density peak of AGN
increases with the luminosity, from $z$$\sim$0.5 at \lx$\sim$10$^{42}$
\ers\ up to $z$$\sim$2.5 at \lx$\sim$10$^{46}$ \ers.  Out of the four
proposed \nh\ distributions only fit \# 4, provides a good fit to the
whole data in the \lx-$z$-\nh\ space.  The first model (fit \# 1)
searched for a constant value of the fraction of objects with
\nh$<$10$^{21}$ \cmm\ ($\Psi(L_{\rm X},z)=\psi=$ constant,
$\beta_L$=$\beta_z$=0).  The $\chi^2$ probabilities of the dependence
of the \nh\ distributions on \lx\ and $z$ reject, at more than 99.93\%
confidence level, this model.  As can be seen in Figures
\ref{fig_NHlzLDDEs} and \ref{fig_NHLDDEs}, the data requires a
decrease of the fraction of absorbed objects with luminosity, and an
increase with redshift.  Both \nh\ distributions in which we allowed
for a dependence of the absorbed objects on redshift or luminosity
only (fits \#2 and \#3) are rejected at more than 99.5\% confidence
level.  On the contrary, model \#4 (Figures
\ref{fig_xlfLDDEslz} and
\ref{fig_lfLDDEslz}), where both a dependence
on redshift and luminosity is allowed (see Figures
\ref{fig_NHLDDEslz} and \ref{fig_FracLDDEslz}), provides a very good
representation of the data with a $\chi^2_{N_H}$ probability of 83\%.

%{\includegraphics[]{Fig_HistNH_SLZ.eps}}

%{\includegraphics[angle=-90]{Fig_Frac_LDDEslz.eps}}

\subsubsection{Analysis of the uncertainties and systematic biases}

We analyzed how much our results could be affected by uncertainties in
the completeness correction method used. These uncertainties could be
introduced by errors in the spectroscopic classification of the AGN,
and by the assumption that the measured fraction of AGN1 as a function
of \lx\ and N$_{H0}$ and the derived two \lx --$L_{\rm R}$
relationships for AGN1 and AGN2 (see \S 3.3) hold also for the higher
redshift, optically fainter, unidentified population. In order to
measure the maximum allowed range of the HXLF parameters due to
uncertainties in these assumptions, we have carried out the HXLF fits
under the two very extreme hypotheses that all the unidentified AGN
would follow either the \lx --$L_{\rm R}$ relationship typical of the
AGN1 (eq. \ref{eq_slpAGN1}), or the \lx --$L_{\rm R}$ relationship
typical of the AGN2 (eq.  \ref{eq_slpAGN2}). It resulted that the
best-fit parameters changed within the measured 1$\sigma$
uncertainties. The results are shown in Figure \ref{fig_xlfLDDEslz},
where the largest allowed AGN density regions due to the uncertainties
introduced by the completeness correction method used are shown by
dashed areas.
  
About 60\% of the AGN used in our analysis have their \nh\ column
densities derived from the hardness ratios (those belonging to the
AMSSn, H2XMM0.5, CDF-N and CDF-S samples).  This method could
introduce some systematic bias. For example, our simple absorbed power
law model could tendentially underestimate the real column densities,
because scattered X--rays and circum--nuclear starburst X--rays can
provide additional flux. This effect is expected to be stronger at
lower luminosities were the fraction of the light coming directly from
the nucleus should be smaller. If this is the case, the observed
decrease of the fraction of absorbed AGN with the intrinsic luminosity
should be even stronger. It should be noted, however, that Perola
\etal\ (2004) found a rather satisfactory correlation between the
column densities measured from the hardness ratios and from the X--ray
spectral fits in the HELLAS2XMM sample.
  
Recently Tozzi \etal\ (2005) have published \nh\ measurements on the
CDF-S sample, obtained using X--ray spectral fits. We took advantage
of these measures to check whether the hardness ratio method
introduces some relevant systematic bias. No relevant difference or
systematic trend on either luminosity or redshift was found.  In a
subsample of $z\leq$1.2 AGN, using the hardness ratios we measure a
fraction of 17/32 absorbed AGN with \lx\ $>$10$^{43}$ \ers , while
Tozzi \etal\ (2005) find 16/32. At lower luminosities (10$^{41}$$<$\lx
$\leq$10$^{43}$ \ers) we measure a fraction of 20/31 absorbed AGN
(\lx\ $>$10$^{43}$), while Tozzi \etal\ (2005) find 18/31.  In a
subsample with 10$^{43}$$<$\lx $\leq$10$^{45}$ \ers at redshift below
1.5 we measure a fraction 23/43 absorbed AGN, while Tozzi \etal\ 
(2005) find 24/43. At redshift above 1.5 we find a fraction 31/43
absorbed AGN, while Tozzi \etal\ (2005) find 34/43.

We also checked whether our results might depend on the assumed X--ray
K-correction (see \S 3.2). We repeated the fit \# 4 assuming
$\Gamma=1.7$ or $\Gamma=1.9$, or assuming an exponential cutoff at
energy $E_C=300$ keV.  It turned out that the changes of the
parameters are within the 1$\sigma$ uncertainties.

%{\includegraphics[]{xlfPLEslz.eps}}
\begin{figure}
\centering
\resizebox{\hsize}{!}
{\includegraphics[]{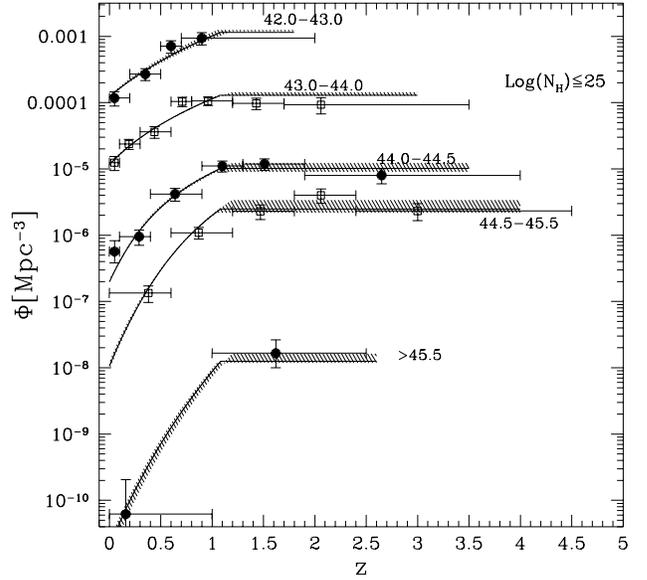}}
\caption{
  Density of AGN in luminosity bins as a function of redshift.  The
  dashed lines are the best fit values of the PLE model with an
  evolving \nh\ distribution (fit \#7 in Table \ref{tab_fit}).  Data
  have been plotted using the ``$N^{\rm obs}/N^{\rm mdl}$ method''
  (see \S4.1). The dashed areas
    are the largest allowed regions due to uncertainties in the
    completeness correction method used (see \S 4.1.1).
}
\label{fig_xlfPLEslz}
\end{figure}

\subsection{PLE model}

We also checked if a simpler pure luminosity evolution model were
consistent with the data. By introducing the evolution factor

\begin{equation}
e(z) = \left\{ \begin{array}{ll}
   (1+z)^{p1}  &(z\leq z_{\rm c})\\
    e(z_{\rm c}) &(z > z_{\rm c}),
\end{array}
\right. \
\label{eq-evolvple}
\end{equation}

\noindent
the PLE model is expressed as

\begin{equation}
\frac{{\rm d} \Phi (L_{\rm X}, z)}{{\rm d Log} L_{\rm X}} 
= \frac{{\rm d} \Phi (L_{\rm X}/e(z), 0)}{{\rm d Log} L_{\rm X}}.
\label{eq-PLE}
\end{equation}

The PLE fit (Figure \ref{fig_xlfPLEslz} and fit \# 6 in Table
\ref{tab_fit}) provides a less probable solution for the HXLF.
Furthermore the PLE fit finds that the evolution stops at
$z_c$$=$1.08$^{+0.08}_{-0.06}$.  This low value should be attributed
to the fact that there is an increase with luminosity of the redshift
peak of the density of AGN.  Low luminosity (\lx$<$10$^{43}$ \ers)
AGNs peak at $z=0.5$, while high luminosity (\lx$>$10$^{46}$ \ers)
AGNs peak at $z\sim2$.  In this framework the PLE fit finds a {\it
  weighted mean} of the different redshift cut off values of the low
and high luminosity AGNs.

Although formally acceptable, $z_c$$=$1.08$^{+0.08}_{-0.06}$ is
significantly smaller than the previous estimates for the evolution of
AGN1 in the hard X--rays ($z_{c}$=$2.4\pm0.5$; La Franca \etal\ 2002),
and in the optical band ($z_{c}$$\sim$2.0; see e.g.  Boyle \etal\ 
2000).  This difference should be attributed to the fact that both the
hard X--ray AGN1 and the optical QSO populate preferentially the bright
part of the HXLF (see e.g.  Figure \ref{fig_frAGN1} and related
discussion) which, also in the LDDE model, faces a redshift cut off
larger than 1.5--2.  If the fit of the PLE model is carried out with a
fixed $z_c$$=$2.0, it turns out unacceptable, with a $\chi^2_{LF}$
probability of 3.4$\times$10$^{-6}$\% and a 2D--KS probability of
0.27\%. On the basis of these results and on the fact that
the PLE model over-predicts the 2--10 keV X--ray background (XRB, see
Table 2) and the soft X--ray counts (see \S 4.3), we consider such a
parameterization of the HXLF evolution to be ruled out.

\subsection{The Counts}

%{\includegraphics[]{Fig_ContHardDiff.eps}}
\begin{figure}
\centering
\resizebox{\hsize}{!}
{\includegraphics[]{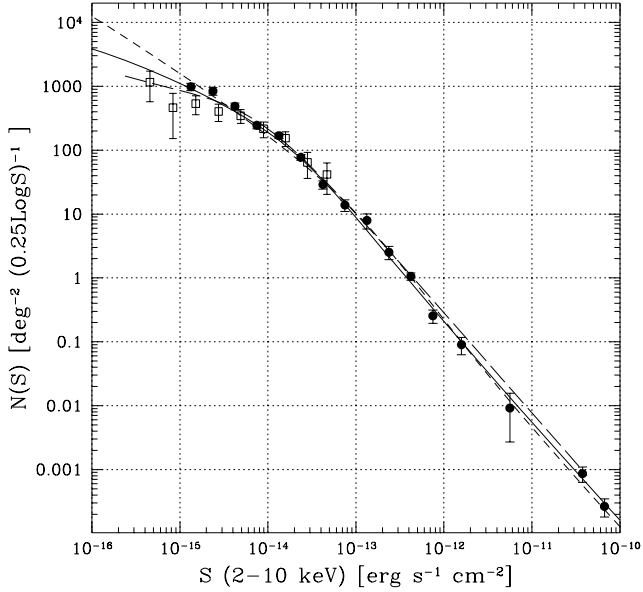}}
\caption{
  Differential counts of AGN in the 2--10 keV band. The filled circles
  are our estimates from the samples used in this analysis.  The open
  squares are the estimates from Bauer \etal\ (2004).  The continuous
  line are the counts predicted by the LDDE model (fit \# 4), while
  the short dashed line are the counts of the PLE model (fit \# 6).
  The long dashed line are the estimates from a compilation of Moretti
  \etal\ (2003).}
\label{fig_CountsH}
\end{figure}

%{\includegraphics[]{CountDiffMany2.eps}}
\begin{figure}
\centering
\resizebox{\hsize}{!}
{\includegraphics[]{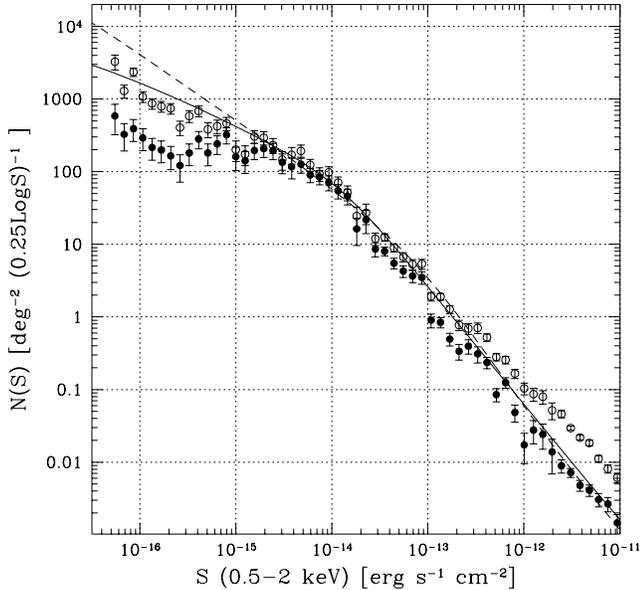}}
\caption{
  Differential counts of X--ray sources in the 0.5--2 keV band (open
  squares) and of AGN-1 (filled squares) from a compilation of
  Hasinger \etal\ (2005).  The continuous line are the
  counts predicted by the LDDE model (fit \# 4), while the dashed line
  are the counts of the PLE model (fit \# 6).  }
\label{fig_CountsS}
\end{figure}

%{\includegraphics[]{Counts4lines.eps}}
\begin{figure}
%\centering
\resizebox{\hsize}{!}
{\includegraphics[]{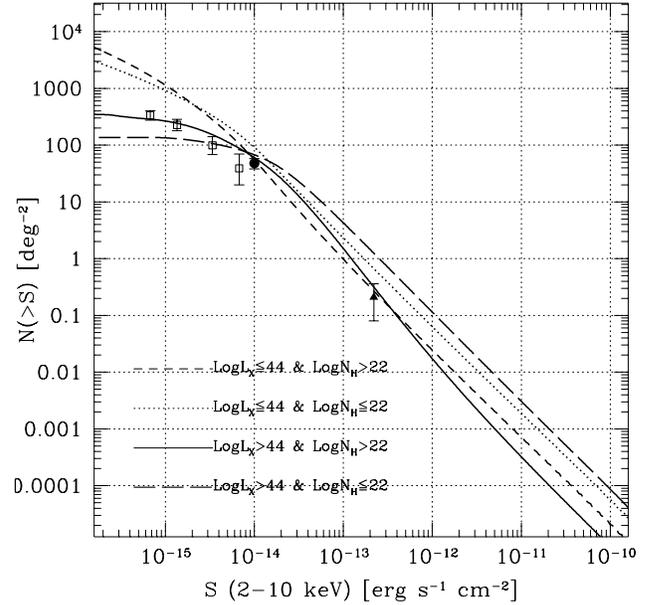}}
\caption{ Predicted counts (from our
  best-fit LDDE model \#4) of AGN for un--absorbed and absorbed AGN
  divided into two luminosity classes. The filled dot is the measure
  of the density of QSO2 by Perola \etal\ (2004), the open circles are
  the density of QSO2 derived by Padovani \etal\ (2004), while the
  triangle is the measure of the density of QSO2 from the HBS28 sample
  of Caccianiga \etal\ (2004).}
\label{fig_Counts4}
\end{figure}

Down to the flux limit adopted in this analysis (10$^{-15}$
\ecs), the 2--10 keV counts predicted by the models described in Table
\ref{tab_fit} are in good agreement with both the counts of the whole
sample (including the objects without spectroscopic identifications),
and with the Bauer \etal\ (2004) compilation (see Figure
\ref{fig_CountsH}).  The fit of Moretti \etal\ (2003) is also shown.
This is an {\it a posteriori} test implying that our method, used to
correct for the spectroscopic incompleteness of the faint samples, is
reliable.  At faint fluxes ($S$$<$10$^{-15}$ \ecs, where there are no
data in our samples) the LDDE model is consistent within the errors
with the data, while the PLE model tends to over-predict the measured
number density \footnote{As explained in \S 3.4, we have included in
  the \nh\ distribution a fraction of objects with 10$^{25}$$<$\nh
  $\leq$10$^{26}$ \cmm , equal to that in the interval 10$^{24}$$<$\nh
  $\leq$10$^{25}$.}. This is mainly due to a higher density of low
luminosity AGNs in the HXLF (in comparison with the LDDE model), and to
the absence of a counter-evolution at high redshift in the PLE model.

Although this analysis is based on observations made in the 2--10 keV
band, it is instructive to compare our results with the 0.5--2 keV
counts. Of course, we should be aware that, when predicting the 0.5--2
keV counts, our results depend on the spectral assumptions (a
$\Gamma$=1.8 spectral slope plus photoelectric absorption), which
could not be valid below 2 keV. The PLE model over-predicts the
observed soft counts as compiled by Hasinger \etal\ (2005) at faint
($S$$<$10$^{-15}$ \ecs) fluxes (see Figure \ref{fig_CountsS}).  The
situation is even worse, since at faint fluxes we expect a relevant
contribution from normal X--ray galaxies to the counts (about 20
deg$^{-2}$ and 400 deg$^{-2}$ at $S_{0.5-2}$=10$^{-15}$ \ecs and
$S_{0.5-2}$=10$^{-16}$ \ecs\ respectively; Ranalli \etal\ 2003, Bauer
\etal\ 2004).  On the contrary, the LDDE model provides a more
acceptable solution. At bright fluxes the observed counts are above
our predictions because the X--ray sources are dominated by stars and
clusters of galaxies which are not included in our models.

\subsubsection{The density of absorbed AGN}

The 2--10 keV predicted counts obtained from the best fit model for
the HXLF (\# 4) are shown in Figure \ref{fig_Counts4}, after being
splitted according to X--ray absorption and X--ray luminosity.  Most
of the luminous (\lx$>$10$^{44}$ erg s$^{-1}$), absorbed (\nh
$>$10$^{22}$ \cmm) sources are AGN2 (see also the discussion in \S
3.3).  Luminous, obscured AGN are usually referred to as QSO2 and in
the simplest version of the AGN unified scheme are predicted to be
more numerous than QSO1 by a factor comparable to that observed for
lower luminosity Seyfert galaxies (about 3--4).  Despite extensive
searches, narrow-line optically luminous QSO2 appear to be extremely
rare and by far less numerous than broad line quasars (see Halpern,
Eracleous \& Forster 1998 and references therein).  Because of the
selection effects due to obscuration, X--ray surveys are expected to
provide an unbiased census of the QSO2 population and a more reliable
estimate of their space density.  Several QSO2 candidates (i.e.
luminous, X--ray obscured sources) have been discovered by {\it
  Chandra} and XMM-{\it Newton} surveys ($\sim20$ in the HELLAS2XMM
survey, Fiore et al. 2003, Mignoli et al. 2004, Cocchia et al.  2005;
$\sim30$ in the CDF-S+CDF-N; a dozen in the CLASXS survey, Barger et
al.  2005).  A sizable fraction of them (from 50\% to 75\%) has been
confirmed by deep optical spectroscopic observations.  A notable
example has been reported by Norman \etal\ (2002).  It should be noted
that the quality of spectroscopic observations is not uniform and,
given the relatively high redshifts, the $H\alpha$ and H$\beta$
wavelengths are poorly sampled, thus hampering a ``pure" optical
classification. On the other hand, it is important to remind that the
QSO2 classification is wavelength dependent.  Several, X--ray
obscured, luminous QSO2 do not show any evidence of strong emission
lines even in high quality optical spectra, among them the QSO2
prototype NGC 6240 (Vignati et al. 1999). Keeping in mind these
caveats, and adopting an admittedly arbitrary luminosity threshold
(\lx$>$10$^{44}$ erg s$^{-1}$), we obtain a QSO2 space density of 60
and 267 deg$^{-2}$ at 2--10 keV fluxes brighter than 10$^{-14}$ and
10$^{-15}$ \ecs respectively.  The slope of the integral counts
significantly flattens below 10$^{-14}$ \ecs. As a consequence, the
QSO2 surface density at fluxes much fainter than those actually probed
(i.e.  10$^{-16}$ \ecs) increases by a relatively small amount
reaching 354 deg$^{-2}$.  These figures have at least a 5\% error,
corresponding to the HXLF normalization uncertainties. The predicted
counts (Figure \ref{fig_Counts4}) are in good agreement with the QSO2
space densities measured by Caccianiga \etal\ (2004), Perola \etal\ 
(2004) and Padovani \etal\ (2004).

\subsection{The XRB spectrum: a self-consistency check}

%{\includegraphics[]{fabio2.ps}}
\begin{figure}
%\centering
\resizebox{\hsize}{!}
{\includegraphics[]{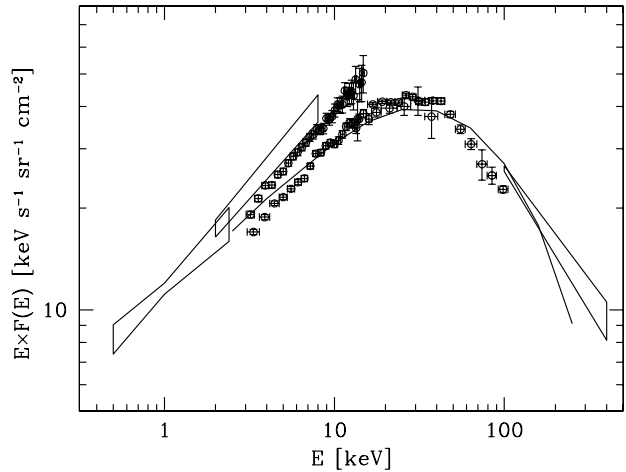}}
\caption{
  Integrated AGN spectrum computed from the best fit LDDE model for
  the HXLF (fit \# 4) with the redshift and luminosity dependent \nh\ 
  function. The model predictions are compared
  with a selection of XRB spectral measurements over the broad
  0.5--400 keV energy range.  The regions enclosed within the
  bow--ties correspond to the XRB spectrum and associated errors as
  measured by ROSAT (Georgantopoulos et al. 1996) in the 0.5--2 keV
  band and by XMM-{\it Newton} (De Luca \& Molendi 2004) in the 2--8 keV band.
  The data points in the 3--15 keV energy range are from {\it RXTE}
  (Revnitsev et al. 2003), while those in the 3--60 keV are from
  {\it HEAO1}--A2 (Marshall et al. 1980). The bow--tie at high energies
  (100--400 keV) is from {\it HEAO1}--A4 (Kinzer et al.  1997).  }
%\figurenum{1}
%\addtolength{\baselineskip}{10pt}
\label{fig_bckgx}
\end{figure}

%{\includegraphics[]{LFAss_NoAss2.eps}}
\begin{figure}
%\centering
\resizebox{\hsize}{!}
{\includegraphics[]{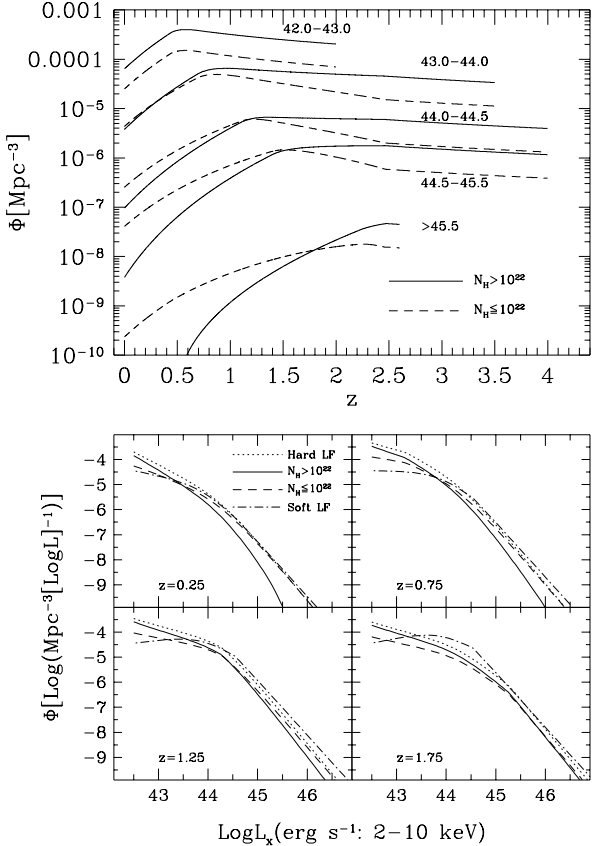}}
\caption{{\it Top.} Density of absorbed and un-absorbed AGN in
  luminosity bins as a function of redshift (fit \# 4).  {\it Bottom.} Density of
  absorbed and un-absorbed AGN as a function of luminosity in four
  redshift intervals (fit \# 4). The dot-dashed line is the 0.5--2 keV LF of AGN
  of Miyaji \etal\ (2000; plotted assuming
  $\Gamma$$=$1.8).
 }
\label{fig_AssNoAss}
\end{figure}

%{\includegraphics[]{PlotLFAGN12.eps}}
\begin{figure}
%\centering
\resizebox{\hsize}{!}
{\includegraphics[]{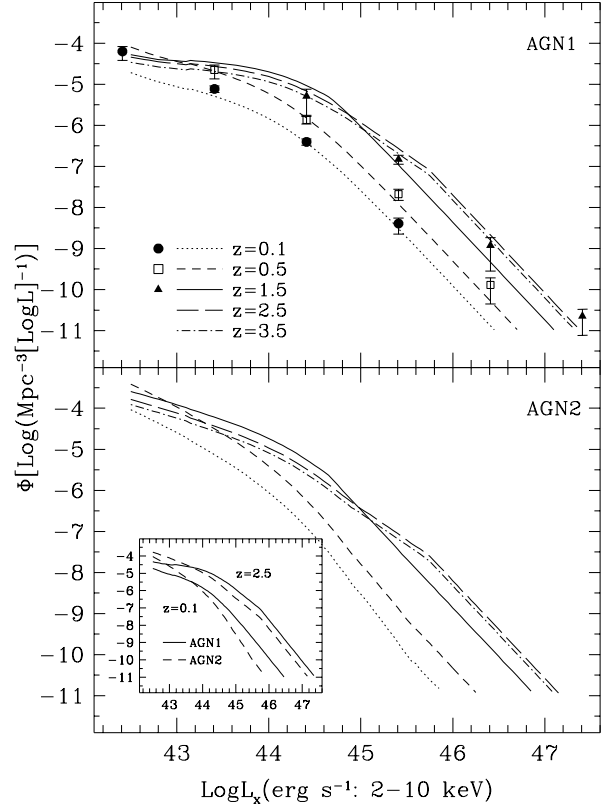}}
\caption{
  {\it Top.}  Evolution of the luminosity function of AGN1 up to
  $z=3.5$. The data are the estimates of the density of AGN1 from La
  Franca \etal\ (2002). {\it Bottom.} Evolution of the luminosity
  function of AGN2 up to $z=3.5$. The lines have the same meaning as
  in the top panel. The small picture shows the difference among the
  luminosity functions of AGN1 and AGN2 at $z=0.1$ and $z=2.5$.  }
\label{fig_LFAGN12}
\end{figure}

A detailed modeling of the XRB spectrum over the full $\sim$2--400 keV
range is beyond the scope of this paper.  Nevertheless, it is
important to check that the evolving X--ray luminosity function and
the $N_H$ distribution derived in the previous sections match the XRB
flux at least in the 2--10 keV energy range.  To this end it is useful
to remind that the XRB intensity below $\sim$10 keV, as measured by
several imaging X--ray telescopes, is likely to be affected by
systematic errors. In Figure \ref{fig_bckgx} a compilation of XRB
measurements is reported. The maximum difference is of the order of
30\% between the {\tt EPIC}--{\it pn} flux reported by De Luca \&
Molendi (2004) and the {\it HEAO1}--A2 measure of Marshall \etal\ 
(1980). According to a recent reanalysis of the {\it HEAO1}--A2 data
(Revnivtsev \etal\ 2004) the 3--60 keV spectrum should be renormalized
upward by about 15\%. The resulting 2--10 keV flux is 1.96$\pm$0.10
\ecs deg$^{-2}$.  The solid curve in Figure \ref{fig_bckgx} represents
the integrated AGN spectrum obtained with our best fit LDDE model for
the HXLF (fit \# 4) with the redshift and luminosity dependent \nh\ 
function\footnote{As explained in \S 3.4, we have included in the \nh\ 
  distribution a fraction of objects with 10$^{25}$$<$\nh
  $\leq$10$^{26}$ \cmm , equal to that in the interval 10$^{24}$$<$\nh
  $\leq$10$^{25}$.}. Our predicted 2--10 keV flux of
1.81$\times$10$^{-11}$ \ecs deg$^{-2}$ corresponds to $\sim$92\% of
the Revnivtsev \etal\ (2004) value and 108\% of the original {\it
  HEAO1}--A2 measure.  Given that the XRB synthesis has been obtained
with very simple prescriptions for the intrinsic (before absorption)
spectral energy distribution (a power law spectrum with $\Gamma$=1.8
plus an exponential high energy cut--off $e^{-E/E_C}$ with $E_C$= 200
keV for all AGN) it is reassuring to obtain a reasonably good
description of the XRB spectral intensity over a broad energy range.
As a final remark we note that an increasing ratio between absorbed
and un--absorbed AGN towards high redshifts has been already included
in the synthesis models of Pompilio \etal\ (2000) and Gilli \etal\ 
(2001) though with different prescription for the absorption
distribution.

\subsection{The LF of absorbed and un-absorbed AGN}

It is interesting to plot the evolution of absorbed (\nh $>$10$^{22}$
\cmm) and un-absorbed AGN, according to our best fit LDDE solution
(fit \# 4; Figure \ref{fig_AssNoAss}). As expected, the absorbed AGN
outnumber the un-absorbed ones at low luminosities and high redshifts.
In the bottom panel of Figure \ref{fig_AssNoAss} the HXLF is compared
with the estimate of Miyaji \etal\ (2000) of the soft X--ray (0.5--2
keV) AGN LF (a slope $\Gamma$=1.8 has been assumed to convert the
0.5--2 keV luminosities into the 2--10 keV band). It turns out that,
at low redshifts ($z$$\sim$0.25), the soft X--ray LF is almost
coincident with our measure of un--absorbed AGN HXLF, while at high
redshifts the soft X--ray LF is consistent, within the uncertainties,
with the total HXLF.  This behavior is explained by the stronger
effects of absorption in the soft X--rays in comparison to hard X--rays,
especially at low redshifts.  As a consequence, at low redshifts, only
un--absorbed AGN are detected in the soft X--ray band, while at high
redshifts the bias reduces, and the same population which is
observed in the 2--10 keV band is detected.  As a consequence, the
soft X--ray LF faces a stronger (LDDE) evolution than observed for the
HXLF (see Hasinger et al. 2005).

\subsection{The LF of AGN1 and AGN2}

The space density and evolution of AGN1 and AGN2 can be estimated
using the above described method.  In order to correct for the
spectroscopic incompleteness of the faint samples we had to compute
the {\it completeness function} $g(L_{\rm X},z,N_H, R)$ (see \S 3.3)
which is based on the estimate of the probability of an AGN to appear
as an AGN1 as a function of \lx , \nh\ and $z$: Q1$(L_{\rm X}, z,
N_{\rm H})$ (shown in Figure \ref{fig_frAGN1}).  With this estimate in
hand (and keeping in mind the uncertainties on the AGN1-AGN2
  optical classification discussed in \S 3.3), we can derive the
AGN1 luminosity function:

\begin{equation}
\Phi_1 (L_{\rm X}, z) 
=\int\Phi (L_{\rm X}, z)
f(L_{\rm X},z;N_{\rm H}) Q1(L_{\rm X},N_{\rm H},z){\rm d}N_{\rm
H}, 
\label{eq_fiAGN1}
\end{equation}

\noindent
where, as discussed in \S 3.4, $f(L_{\rm X},z;N_{\rm H})$ is the \nh\ 
distribution.  The AGN2 density can be derived by substituting, in the
above formula (eq. \ref{eq_fiAGN1}), $Q1(L_{\rm X}, N_{\rm H}, z)$
with 1-$Q1(L_{\rm X}, N_{\rm H}, z)$.  As can be seen in Figure
\ref{fig_LFAGN12}, at low redshifts ($z$$<$0.5), the AGN2 density low
luminosities (\lx$\sim$10$^{42}$--10$^{43}$\ers) is about five times
larger than that of AGN1, while the latter outnumber the former by an
order of magnitude at high luminosities (\lx$\sim$10$^{46}$). In
Figure \ref{fig_LFAGN12} the AGN1 density in the 2--10 keV band from
the {\it BeppoSAX} HELLAS survey as computed by La Franca \etal\ 
(2002) is reported.  The present estimate of the AGN1 luminosity
function is consistent, within the uncertainties, with the La Franca
\etal\ (2002) findings, thus confirming that Q1 is a reliable measure
of the probability of an AGN to appear as an AGN1.

%%%%%%%%%%%%%%%%%%%%%%%%%%%%%%%%%%%%%%%%%%%%%%%%%%%%%%%%%%%%%%%%%%%%%%%%%%%%
\section{Discussion}

\subsection{Comparison with previous results}

A specific procedure to correct for the spectroscopic incompleteness
of faint X--ray sources which also takes into account the selection
effects due to X--ray absorption has allowed us to use a large AGN
sample to compute the HXLF.  Our results extend those of Cowie \etal\ 
(2003) and Barger \etal\ (2005), where no correction for X--ray
absorption is adopted, and the upper limits to the AGN density are
estimated by assigning to the unidentified sources the redshifts
corresponding to the centers of each \lx-$z$ bin.

A LDDE model provides the best fit to the HXLF evolution up to $z=4$,
in agreement with the Ueda et al. (2003) findings obtained using a
smaller and brighter sample, and also with the estimates of Cowie
\etal\ (2003), Fiore \etal\ (2003), Hasinger \etal\ (2005) and
Silverman \etal\ (2005) who found that the AGN number density for
luminosities lower than $\sim 10^{44}$ \ers peaks at lower redshifts
than that of higher luminosity AGN.

%\plotone{Fig_Frac.eps}
\begin{figure}
\centering
\epsscale{1.15}
\plotone{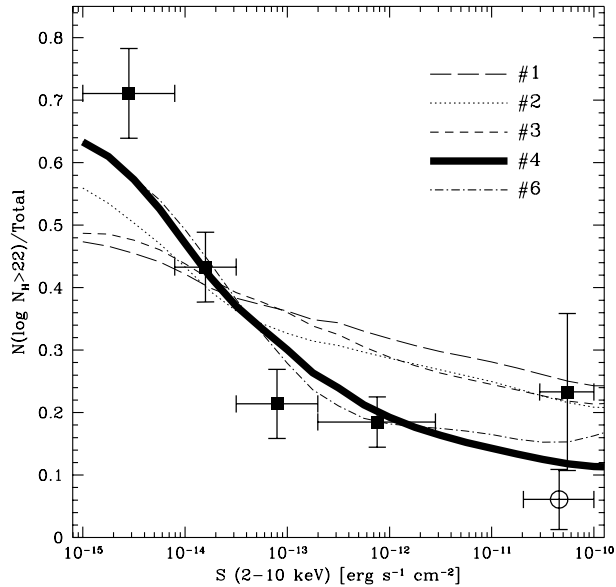}
\caption{Observed fraction of absorbed AGNs as measured from the samples
used in this analysis. The lines correspond to the predictions of the
fitted HXLF listed in Table \ref{tab_fit}. The open circle is the
value obtained from the sample of Grossan (1992; not included in the analysis
of the HXLF).
}
\label{fig_frflx}
\end{figure}

The new result or our analysis concerns the luminosity and redshift
dependence of the fraction of absorbed AGN, which decreases with
luminosity and increases with redshift.  The luminosity trend was
already pointed out by Ueda \etal\ (2003) (see also Hasinger et al.
2005).  In the Ueda \etal\ (2003) best fit model the fraction of
absorbed AGN is 57\% and 36\% at \lx=10$^{42.5}$ \ers and
\lx=10$^{45}$ \ers respectively.  Taking into account only a
luminosity dependence on the fraction of absorbed AGN in our sample
(fit \# 2 in Table 2), the corresponding fractions at \lx=10$^{42.5}$
\ers and \lx=10$^{45}$ \ers\ are 68\% and 40\% respectively. The two
results are remarkably similar, especially if we note that absorbed AGN
with 24$<$Log\nh$<$25 are included in our sample but not in Ueda et
al. (2003).
 
The increase of the fraction of absorbed AGN with redshift,
instead, emerges only with our analysis.  The difference with respect
to the Ueda et al. (2003) findings is due to the larger sample
extending to fainter fluxes used in the present analysis.  Indeed, if
we restrict our analysis to a subsample (fit \#5 in Table 2) of 207
objects from the Piccinotti, AMSSn and CDF-N catalogs and thus
quantitatively similar to that used by Ueda et al. (2003) the
uncertainties become so large that the redshift dependence is no
longer significant while the luminosity dependence is recovered.

%(Figure \ref{fig_NHlzLDDEslz}).

It is worth noting that the luminosity and redshift dependence of
the absorbed AGN fraction would disappear if one flux limited sample
only were analyzed (as discussed by Perola \etal\ 2004).  A flux
limited sample selects low luminosity AGN at low redshifts (which,
according to our analysis, are more absorbed) and high luminosity AGN
at high redshift (which are more absorbed as well!).  Then the average
fraction of absorbed AGN turns out to be roughly constant. Only
combining several samples, and thus covering wide strips of the \lx
--$z$ plane with almost constant redshift or luminosity, it is possible
to disentangle the true dependencies.

A simple AGN model based on the unified paradigm has been adopted by
Treister \etal\ (2004). Assuming that obscured AGN outnumber
unobscured ones by a factor 3 without any luminosity and/or redshift
dependence they claim to be able to reproduce the observed counts,
redshift and \nh\ distributions in the CDF-N and CDF-S samples once
all the selection effects are properly taken into account. More
recently Treister \& Urry (2005) revised their previous analysis
including a luminosity dependence of the fraction of absorbed AGN
which appears to provide an equally good fit to several observational
constraints. However, in both works, no comparison between the
predicted and observed \nh\ distributions as a function of both \lx\ 
and $z$ is made.  We have repeated our analysis assuming the Treister
\& Urry (2005) \nh\ distribution (see their Fig. 1). Either using the
CDF-N plus CDF-S samples only, or the full AGN sample used in this
work, we found that the only statistically acceptable models are those
including a dependence of the fraction of absorbed AGN as a function
of \lx\ and $z$, with a behaviour similar to what measured in the
present paper (see \S 4).

The observed and predicted fractions of absorbed
(\nh$>$10$^{22}$/Total) AGN as a function of the observed flux are
shown in Figure \ref{fig_frflx}.  The open circle is from the Grossan
sample (1992; not included in this analysis\footnote{A proper reassesment
of the Grossan sample seems necessary before using it extensively for a 
detailed statistical analysis. See e.g. Bianchi et al. (2005) for a discussion
on a few sources of the sample.}), and it is plotted in
order to show the uncertainties at bright fluxes.  As already
described by several authors (e.g. Comastri \etal\ 2001; Tozzi \etal\ 
2005; Perola \etal\ 2004 and references therein) the average X--ray
spectrum significantly hardens towards faint fluxes and this change is
mostly concentrated in the 10$^{-14}$--10$^{-13}$ \ecs range, where the
fraction of absorbed AGNs rises from about 20\% to about 50\%. For
this reason, this measure is a very powerful tool to discriminate
between different evolutionary scenarios for the \nh\ distribution.
The only acceptable description of the observed ratio between absorbed
and un--absorbed AGN as a function of the hard X--ray flux is obtained
only if the ratio depends on {\it both} luminosity {\it and} redshift
(fits \#4 and \#6 in table 2 for LDDE and PLE, respectively).
These two models are indistinguishable, and in fact the fraction of
absorbed AGN as a function of the X--ray flux is less sensitive to the
shape and evolution of the HXLF than to the evolution of the \nh\ 
distribution (see also \S 3.5).

Recently, Alexander \etal\ (2005a, 2005b) have found evidences that a
fraction of the $z$$>$1 submillimiter emitting galaxies harbor
obscured AGN. They argued that the black holes are almost continuously
growing throughout vigorous star--formation episodes. These results
are in agreement with the hydrodinamical simulation of galaxy mergers
by Di Matteo \etal\ (2005) and Springel \etal\ (2005), where the
growth of both the black holes and stellar components are taken into
account.  In this framework, our result of an increase of the fraction
of absorbed AGN with the redshift is in agreement with a picture where
the peak epoch of the star formation ($z$=1--2) corresponds to a
heavily obscured rapid black-hole phase, which is ultimately proceeded
by an unobscured quasar phase (Alexander \etal\ 2005a, 2005b, Hopkins
\etal\ 2005).

\subsection{Accretion history of the Universe}

%{\includegraphics[]{Accretion.eps}}
\begin{figure}
\centering
\resizebox{\hsize}{!}
{\includegraphics[]{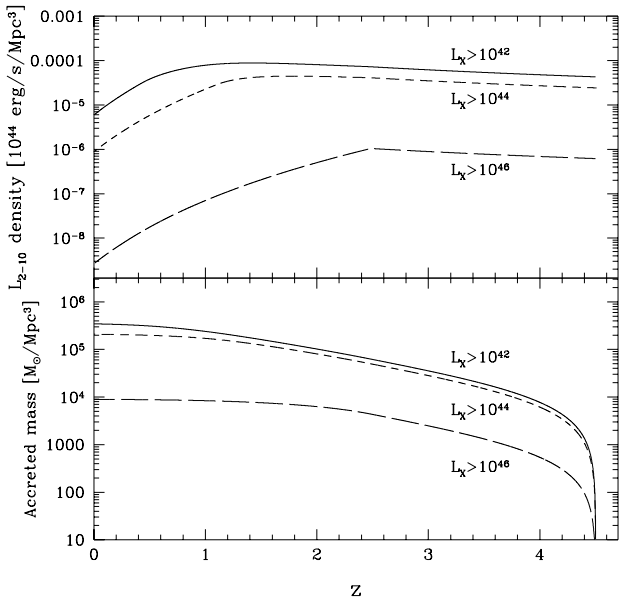}}
\caption{{\it Top.} Intrinsic (before absorption) luminosity density in the 2--10 keV
band as a function of redshift from our best fit HXLF. 
{\it Bottom.} Total accreted mass as a function of redshift.
 }
%\figurenum{1}
%\addtolength{\baselineskip}{10pt}
\label{fig_accretion}
\end{figure}

Our measure of the HXLF cosmological evolution directly constrains the
history of the formation of supermassive black holes (SMBH) in the
galactic centers not only for luminous un-obscured AGN, that can be
traced also at longer wavelengths (optical, soft X--rays), but also for
the less luminous or obscured AGN. Starting from our best fit HXLF it
is possible to derive the {\it intrinsic} (i.e. before absorption)
luminosity density in the 2--10 keV band in the Universe as a function
of redshift:

\begin{equation}
\int L_{\rm X} \Phi (L_{\rm X}, z){\rm d Log} L_{\rm X}.
\end{equation}
 
This quantity can be converted into the energy density production rate
per comoving volume by means of a bolometric correction
factor $K$ ($L_{bol} = K L_{\rm X}$).  The mass inflow rate onto a
SMBH, $\dot M_\bullet$, is related to the bolometric luminosity of the
AGN, $L_{bol}$, by $\dot{M}_\bullet = L_{bol}(1-\epsilon)/\epsilon
c^{2}$, where $\epsilon$ is the radiative efficiency of 
the accretion flow (typically taken
to be about 0.1; see e.g. Yu \& Tremaine 2002, Marconi \etal\ 2004,
Barger \etal\ 2005). Once  a value for $\epsilon$ and $K$ is adopted 
it is straightforward  to derive the accretion rate density as a function of
redshift:

\begin{equation}
\dot{\rho}_{\bullet}(z)= \frac{1-\epsilon}{\epsilon c^{2}} 
\int K L_{\rm X} \Phi (L_{\rm X}, z){\rm d Log} L_{\rm X},
\end{equation}

\noindent
and the total accreted mass, i.e. the total density in massive black
holes, if we assume that the initial mass of seeds black holes  at redshift
$z_s$ is negligible with respect to the total mass:

\begin{equation}
\rho_{BH}(z) = \int_z^{z_s} \dot{\rho}_{\bullet}(z) \frac{{\rm d}t}{{\rm d}z} {\rm d}z. 
\end{equation}

In the top panel of Figure \ref{fig_accretion} we show our direct
estimate of the {\it intrinsic} luminosity density in the 2--10 keV
band as a function of redshift from our best fit HXLF. 
Assuming $\epsilon=0.1$ and the luminosity dependent bolometric 
correction extensively discussed by Marconi et al. (2004; their eq. 21), 
the total density of massive black holes as a function of
redshift reported in Figure \ref{fig_accretion} is obtained 
by integration of the HXLF
starting from $z_s=4.5$, for \lx$>$10$^{41}$ \ers and \nh$<$10$^{26}$
cm$^{-2}$ (as explained in \S 3.4).  
The final accreted mass correspond to a black hole mass
density in the local Universe of $\rho_{BH} =
3.2\ h^2_{70}\times 10^5$ M$_\odot$ Mpc$^{-3}$.
A somewhat higher value $\rho_{BH} = 4.0\ 
h^2_{70}\times 10^5$ M$_\odot$ Mpc$^{-3}$ is obtained 
for a single valued  bolometric correction factor  
$K=40$ (Elvis \etal\ 1994).
These results are consistent, within the
errors, with the SMBH density estimate of 
$\rho_{BH} = 4.6^{+1.9}_{-1.4}\ h^2_{70}\times 10^5$ derived from 
dynamical studies of local galaxies bulges (see e.g. Marconi
\etal\ 2004 and Ferrarese 2002).  As shown in Figure
\ref{fig_accretion}, the vast majority of the accretion rate density
and black hole mass is produced by the low luminosity AGN
(\lx$<$10$^{44}$-10$^{45}$ \ers) down to redshift $z$$\sim$1.  As
already shown by the LDDE model of the HXLF, high luminosity AGN are
already formed at redshift $\sim$2 while low luminosity AGN keep
forming down to $z$$\sim$1. This result is in qualitative agreement
with semi-analytical models for galaxy formation
and star formation rates, such as those of Balland \etal\
(2003), Menci \etal\ (2004), Granato \etal\ (2004), or with the
hydrodinamical simulations such as those of Di Matteo \etal\ (2005),
Springel \etal\ (2005), and Hopkins \etal\ (2005).

%%%%%%%%%%%%%%%%%%%%%%%%%%%%%%%%%%%%%%%%%%%%%%%%%%%%%%%%%%%%%%%%%
\section{Conclusions}

We have devised a method to compute the AGN HXLF which allow us 
to correct for both the spectroscopic incompleteness of the faint
samples, and for the selection effects due to the X--ray K--correction.
Thanks to this method we have been able to collect a sample of about
500 AGN up to $z=$4. The most important results can be summarized as follows:

\begin{itemize}
  
\item{} There is evidence that the fraction of absorbed
  (\nh$>$10$^{22}$ cm$^{-2}$) AGN decreases with the X--ray luminosity,
  and increases with the redshift.

\item{} the AGN HXLF up to $z$=4 is best represented by a LDDE
  model where the low luminosity (\lx$\sim$10$^{43}$ erg s$^{-1}$) AGN
  peak at $z\sim 0.7$ while high luminosity AGN
  (\lx$>$10$^{45}$ erg s$^{-1}$) peak at $z\sim 2$.
  
\item{} We can rule out a PLE model on the basis of several arguments 
 which
  take into account the discrepancies with the optical and hard X--ray
  LF of AGN1, and the over-predictions of the soft X--ray counts 
  and XRB intensity.

\item{} We estimate a density of supermassive black holes in the local
  Universe of $\rho_{BH} = 3.2\ h^2_{70}\times 10^5$ M$_\odot$
  Mpc$^{-3}$, which is consistent with the recent estimates of 
 local galaxies  black hole mass function.

\end{itemize}
%\bullet

\acknowledgements 

The authors are grateful to the referee for helpful comments and
constructive criticism improving the manuscript.  We acknowledge A.
Marconi for useful discussions, G. Hasinger, and A. Moretti for having
provided data in machine readable format, and P. Tozzi for having
provided the \nh\ measurements of the sources in the CDF-S sample
before publication.  This research has been partially supported by
ASI, INAF-PRIN 270/2003 and MIUR Cofin-03-02-23 grants.

%%%%%%%%%%%%%%%%%%%%%%%%%%%%%%%%%%%%%%%%%%%%%%%%%%%%%%%%%%%%%%%

\clearpage

%
%%{\includegraphics[angle=-90]{Fig_Frac_LDDEUeda.eps}}
%\begin{figure}
%\centering
%\resizebox{\hsize}{!}
%{\includegraphics[angle=-90]{f19.eps}}
%\caption{
%  Observed fraction of absorbed (\nh$>$10$^{22}$ \cmm) AGN as a
%  function of \lx\ and $z$ using the Piccinotti, AMSSn and CDF-N
%  samples only.  The dotted lines are examples of the intrinsic
%  assumed distributions at various luminosities and redshifts (LDDE
%  model, fit \#5 in Table \ref{tab_fit}).  The long dashed lines are
%  the corresponding average intrinsic assumed distributions of the
%  sample used.  The short dashed lines are the expectations taking
%  into account the selection effects.  }\label{fig_NHlzLDDEslz}
%\end{figure}
%


\begin{thebibliography}{}


%\bibitem{}

\bibitem[]{} Akiyama, M., Ueda, Y., Ohta, K., Takahashi, T., Yamada, T. \etal\ 2003, \apjs, 148, 275
\bibitem[]{} Alexander, D.M., \etal\ 2003, \aj, 126, 539
\bibitem[]{} Alexander, D.M., \etal\ 2005a, \nat, 434, 738
\bibitem[]{} Alexander, D.M., Bauer, F.E., Chapman, S.C., Smail, I., Blain, A.W., Brandt, W.N., 
Ivison, R.J. 2005b, \apj, in press (astro-ph/0506608)
\bibitem[]{} Anders, E., Grevesse, N., 1989, \gca, 53, 197
\bibitem[]{} Antonucci, R. 1993, \araa, 31, 473
\bibitem[]{} Baldi, A., Molendi, S., Comastri, A., Fiore, F., Matt, G., Vignali, C. 2002, \apj, 564, 190
\bibitem[]{} Balland, C., Devriendt, J.E.G., Silk, J. 2003, \mnras, 343, 107
\bibitem[]{} Barger, A.J.,  et al. 2003, \aj, 126, 632 
\bibitem[]{} Barger, A.J., Cowie, L.L., Mushotzky, R.F., Yang, Y., Wang, W.-H., Steffen, A.T., Capak, P. 2005, \aj, 129, 578  
\bibitem[]{} Bauer, F.E., Alexander, D.M., Brandt, W.N., Schneider, D.P., Treister, E., Hornschemeier, A.E., \&   Garmire, G.P., 2004, AJ, 128, 2048
\bibitem[]{} Bianchi, S., Guainazzi, M., Matt, G., Chiaberge, M., Iwasawa, K., 
Fiore, F.,  Maiolino, R., 2005, A\&A, in press (astro-ph/0507323)
\bibitem[]{} Boyle, B.J., Georgantopoulos, I., Blair, A.J., Stewart, G.C., Griffiths, R.E., Shanks, T., Gunn, K.F., Almaini, O. 1998, \mnras, 296, 1
\bibitem[]{} Boyle, B.J., Shanks, T., Croom, S.M., Smith, R.J., Miller, L., Loaring, N., Heymans, C. 2000, \mnras, 317, 1014
\bibitem{} Brandt, W.N., Hasinger, G 2005, \araa, 43, in press (astro-ph/0501058)
\bibitem{} Brusa, M. \etal\ 2003, \aap, 409, 65
\bibitem{} Caccianiga, A., \etal\ 2004, \aap, 416, 901
\bibitem{} Cocchia et al. 2005, \aap, submitted
\bibitem{} Comastri, A., Setti, G., Zamorani, G., Hasinger, G. 1995, \aap, 296, 1
\bibitem{} Comastri, A., Fiore, F., Vignali, C., Matt, G., Perola, G.C., La Franca, F. 2001, \mnras, 327, 781
\bibitem{} Cowie. L.L., Barger, A.J., Bautz, M.W., Brandt, W.N., Garmire, G.P. 2003, ApJ, 584, L57
\bibitem{} Croom, S.M., Smith, R.J., Boyle, B.J., Shanks, T., Miller, L., Outram, P.J., Loaring, N.S. 2004, \mnras, 349, 1397
\bibitem{} De Luca, A \& Molendi, S. 2004, \aap, 419, 837
\bibitem{} Di Matteo, T., Springel, V., \& Hernquist, L. 2005, \nat, 433, 604
\bibitem{} Elvis, M., et al. 1994, \apjs, 95, 1
\bibitem{} Fasano, G. \& Franceschini, A. 1987, \mnras, 225, 155
\bibitem{} Ferrarese, L., 2002, ApJ, 578, 90 
\bibitem{} Fiore, F., et al. 2000, \na, 5, 143
\bibitem{} Fiore, F., et al. 2003, \aap, 409, 79
\bibitem{} Gehrels, N. 1986, \apj, 303, 336
\bibitem{} Georgantopoulos, I., Stewart, G.C., Shanks, T., Boyle, B.J., Griffiths, R.E. 1996, \mnras, 280, 276
\bibitem{} Giacconi, R., et al.\ 2002, \apjs, 139, 369
\bibitem{} Gilli, R., Risaliti, G., Salvati, M. 1999, \aap, 347, 424
\bibitem{} Gilli, R., Salvati, M., Hasinger, G. 2001, \aap, 366, 407
\bibitem{} Granato, G., De Zotti, G., Silva, L., Bressan, A., Danese, L. 2004, \apj, 600, 580
\bibitem{} Grossan, B.A., 1992, Ph. D. Thesis, MIT, Cambridge
\bibitem{} Halpern, J.P., Eracleous, M., Forster, K. 1998, \apj, 501, 103
\bibitem{} Hasinger, G., 2003, in AIP Conf. Proc. 666, The Emergence of Cosmic Structure, ed. S.S. Holt \& C. reynolds (New York: AIP), 227
\bibitem{} Hasinger, G., Miyaji, T., Schmidt, M. 2005, \aap, in press (astro-ph/0506118)
\bibitem{} Hopkins, P.F., Hernquist, L., Martini, P., Cox, T.J., Robertson, B.,
Di Matteo, T., \& Springel, V. 2005, \apj, in press (astro-ph/0503055)
\bibitem{} Kinzer R.L., Jung G.V., Gruber D.E., Matteson J.L., Peterson L.E. 1997, \apj, 475, 361
\bibitem{} La Franca, F., Franceschini, A., Cristiani, S., \& Vio, R. 1995, \aap, 299, 19
\bibitem{} La Franca, F., Cristiani, S. 1997, \aj, 113, 1517
\bibitem{} La Franca, F., et al. 2002, \apj, 570, 100
\bibitem{} Lampton, M., Margon, B., Bowyer, S. 1976, \apj, 207, 894
\bibitem{} Mainieri, V., Bergeron, J., Hasinger, G., Lehmann, I., Rosati, P., Schmidt, M., Szokoly, G., \& Della Ceca, R. 2002, \aap, 393, 425
\bibitem{} Maiolino, R. \& Rieke, G.H., 1995, \apj, 454, 95
\bibitem{} Maccacaro, T., Della Ceca, R., Gioia, I.M., Morris, S.L., Stocke, J.T., Wolter, A. 1991, \apj, 374, 117
\bibitem{} Marconi, A., Risaliti, G., Gilli, R., Hunt, L.K., Maiolino, R., Salvati, M. 2004, \mnras, 351, 169
\bibitem{} Marshall F.E., Boldt E.A., Holt S.S., Miller R.B., Mushotzky R.F., Rose L.A., Rothschild R.E., Serlemitsos P.J. 1980, \apj, 235, 4
\bibitem{} Menci, N., Fiore, F., Perola, G. C., Cavaliere, A. 2004, \apj, 606, 58
\bibitem{} Mignoli, M. \etal 2004, A\&A, 418, 827
\bibitem{} Miyaji, T., Hasinger, G., Schmidt, M. 2000, \aap, 353, 25
\bibitem{} Moretti, A., Campana, S., Lazzati, D., Tagliaferri, G. 2003, \aap, 403, 297
\bibitem{} Morrison, R., McCammon, D.A., 1983, \apj, 270, 119
\bibitem{} Norman, C. \etal\ 2002, \apj, 571, 718
\bibitem{} Padovani, P., Allen, M.G., Rosati, P., Walton, N.A. 2004, \aap, 424, 545
\bibitem{} Page, M.J., et al. 2003, Astronomische Nachrichten, 324, 101 
\bibitem{} Panessa, F., Bassani, L. 2002, \aap, 394, 435 
\bibitem{} Perola, G.C., et al. 2004, \aap, 421, 491
\bibitem{} Piccinotti, G., Mushotzky, R.F., Boldt, E.A., Holt, S.S., Marshall, F.E., Serlemitsos, P.J., Shafer, R.A. 1982, \apj, 253, 485
\bibitem{} Pompilio, F., La Franca, F., Matt, G., 2000, \aap, 353, 440
\bibitem{} Ranalli, P., Comastri, A., Setti, G. 2003, \aap, 399, 39     
\bibitem{} Revnivtsev, M., Gilfanov, M., Sunyaev, R., Jahoda, K., Markwardt, C. 2003, \aap, 411, 329
\bibitem{} Revnivtsev, M., Gilfanov M., Jahoda, K., Sunyaev, R., 2004, \aap, submitted (astro-ph/0412304)
\bibitem{} Risaliti, G., Maiolino, R., Salvati, M. 1999, \apj, 522, 157
\bibitem{} Setti, G., Woltjer, L., 1989, \aap, 224, L21
\bibitem{} Silverman, J.S.,  et al. 2005, ApJ, 618, 123
\bibitem{} Steffen, A.T., Barger, A.J., Cowie, L.L., Mushotzky, R.F., Yang, Y. 2003, ApJ, 506, L23
\bibitem{} Szokoly, G. P., et al. 2004, \apjs, 155, 271
\bibitem{} Tozzi, P. \etal\ 2005, \aap, submitted
\bibitem{} Treister, E. \etal\ 2004, \apj, 616, 123
\bibitem{} Treister, E. \& Urry, C.M. 2005, \apj, in press (astro-ph/0505300) 
\bibitem{} Ueda, Y., Akiyama, M., Ohta, K., Miyaji, T. 2003, \apj, 598, 886
\bibitem{} Vignati, P. \etal\ 1999, \aap, 349, L57
\bibitem{} Yu, Q., Tremaine, S. 2002, \mnras, 335, 965
\bibitem{} Zheng, W., et al. 2004, \apjs, 155, 73
\bibitem{} Zombeck, M.V. 1990, Handbook of Space Astronomy and Astrophysics
(Cambridge University Press)

\end{thebibliography}
\end{document}